\documentclass[twocolumn,showpacs,preprintnumbers,amsmath,amssymb]{revtex4-1}
\usepackage{epsfig} \usepackage{bm}

\begin{document}

\title{Three-body bremsstrahlung and the rotational character of the
  $^{12}$C-spectrum}

\author{E. Garrido$\:^1$, A.S. Jensen$\:^2$, D.V. Fedorov$\:^2$}
\affiliation{$^1$ Instituto de Estructura de la Materia, IEM-CSIC,
Serrano 123, E-28006 Madrid, Spain}
\affiliation{$^2$ Department of Physics and Astronomy, Aarhus University, 
DK-8000 Aarhus C, Denmark} 

\date{\today}

\begin{abstract}
The electric quadrupole transitions between $0^+$, $2^+$, and $4^+$
states in $^{12}$C are investigated in a $3\alpha$ model. The three-body
wave functions are obtained by means of the hyperspherical adiabatic expansion
method, and the continuum is discretized by imposing a box boundary condition.
Corresponding expressions for the continuum three-body ($3\alpha$)
bremsstrahlung and photon dissociation cross sections are derived and
computed for two different $\alpha-\alpha$ potentials.  The available
experimental energy dependence is reproduced and a series of other
cross sections are predicted.  The transition strengths are defined
and derived from the cross sections, and compared to schematic
rotational model predictions. The computed properties of the $^{12}$C
resonances suggest that the two lowest bands are made, respectively, by
the states $\{0^+_1, 2^+_1, 4^+_2\}$ and $\{0^+_2, 2^+_2, 4^+_1\}$.
The transitions between the states in the first band are consistent
with the rotational pattern corresponding to three alphas in an 
equal sided triangular structure. For the second band, the transitions 
are also consistent with a rotational pattern, but with the three alphas
in an aligned distribution.
\end{abstract}


\maketitle

\section{Introduction}

The structure of the $^{12}$C spectrum has attracted a lot of
attention along the years. In this nucleus only two bound states
exist, the $0_1^+$ ground state and the first excited state, with
angular momentum and parity $2_1^+$ and excitation energy of 4.44 MeV.
We use $J^\pi_i$ to represent the $i^{th}$ state with angular momentum
$J$ and parity $\pi$.  Already in the 50's, F. Hoyle predicted the
existence of a $0_2^+$ resonance at an excitation energy of 7.65 MeV,
as a requirement in order to explain the known abundance of carbon in
the Universe \cite{hoy54}. The existence of such resonance was
experimentally confirmed just a few years later \cite{coo57}.  Its
properties have recently been thoroughly reviewed \cite{fre14}.  Since
then, many other resonant states have been observed in $^{12}$C
\cite{ajz90,fre09,kir10,zim13,zim13b,mar14}.  Among them, one of the
most elusive ones has been the second $2_2^+$ state, which was
predicted by different theoretical methods to have an excitation
energy of around 10 MeV \cite{des87,alv07,che07}.  Recent experiments
have confirmed the existence of this $2_2^+$ resonance at the expected
energy \cite{ito04,fre09,zim13,zim13b}.

Together with the $0^+$ and $2^+$ states mentioned above, a $4^+$
resonance is known to be at an excitation energy of 14.1 MeV
\cite{ajz90,kir10}. Other 4$^+$ states have also been obtained
numerically. In fact, the known $4^+$ resonance is usually
found to be the $4^+_2$ state, and a first resonance
at a lower energy is often obtained (for instance, at 11 MeV in 
\cite{cuo13} or 10.5 MeV in \cite{alv07}). 
Experimental evidence of the $4^+_1$ state was reported in 
Ref.\cite{fre11}, and its excitation energy was given to be
$13.3\pm 0.2$ MeV, only about 1 MeV lower than the $4^+_2$
state. 

The appearance of the $\{0^+,2^+,4^+\}$ sequences in the
energy spectrum suggests the rotational character of these
states. Therefore at least two rotational bands could exist in
$^{12}$C, one of them sitting on the ground state, and another one
sitting on the Hoyle state. In fact, it is becoming very common in the
literature to refer to these states as rotational states
\cite{gai13,ogl14}.

The rotational sequences of states are not necessarily
$\{0^+,2^+,4^+\}$, since the underlying intrinsic shapes different
from axial and $R_2(\pi)$ symmetry may provide additional states in a
given band.  This is the classical knowledge that the quantum numbers
specifying the states in a rotational band directly carry information
about the symmetry of the intrinsic state.  The details of
consequences of the $D_{3h}$ symmetry (equilateral triangle) for the
$3\alpha$ spectrum of $^{12}$C was formulated and discussed in general
in \cite{bij02}.  Evidence for that symmetry was provided in
\cite{mar14} by comparing energy sequences and transition
probabilities.  We emphasize that also this model treats the
resonances as bound states where all strengths are collected in the
corresponding bound-state wave function.

A simple rotational sequence is also observed in $^8$Be, where the
ground state and the two first excited states follow as well the
angular momentum and parity sequence $0^+$, $2^+$, and $4^+$ (also
$6^+$ and $8^+$ states with large widths, are found numerically as
poles of the ${\cal S}$-matrix). All the states in the $^8$Be spectrum
are unbound (resonances), with the ground state only about 0.1 MeV
above the threshold for emission of two alpha particles. Therefore,
the question arises about a possible rotational (or in general
collective) character for a sequence of continuum states of
considerable width.  This problem was recently investigated in
previous works \cite{gar12,gar13,gar14}.  This was done by computation
of the electric quadrupole cross sections for bremsstrahlung emission
after transitions between the $^8$Be-states, which were described as
two-body systems made of two alpha particles.  These calculations
required a careful treatment of the continuum wave functions and
clarification of the definition of the cross section for transitions
between states with a none well-defined energy.

From these $^8$Be cross sections it is possible to extract the
transition strengths, which for the case of transitions within a
rotational band must follow well-established rules based on the
assumption that all the states have the same intrinsic spatial structure 
(rigid rotor).  The main conclusion was that the computed transition
strengths do not behave as expected for states in a rotational
band. In fact, when increasing the angular momentum of the initial
states the transition strength was found to decrease, which is
precisely the opposite to the prediction of the rotational model
\cite{gar13}. Nevertheless, allowing the separation between the two
alphas in the $^8$Be-resonances to change according to the computed
root mean square radius, a very nice agreement between the computed
strengths and the rotational model predictions was observed.

The purpose of the present work is to extend the $^8$Be continuum
investigations to the spectrum of $^{12}$C, and check if the $\{0^+_1,
2^+_1, 4^+_1 \}$ and $\{0^+_2, 2^+_2, 4^+_2 \}$ sequences of states
follow the behavior predicted by the rotational model.  For this aim
we shall compute the electric quadrupole $\gamma$-emission
bremsstrahlung cross sections for the different transitions between
the states. These cross sections and the transition strengths are not 
well-defined quantities when involving
resonances with finite width.  Suitable definitions are necessary to
be specific. We shall work entirely within the strict two- and three-body
framework. This means that the constituent particles are inert
$\alpha$-particles, and the inter-particle nucleonic Pauli-principle
will be accounted for without use of the intrinsic microscopic
structure. This model, used throughout the present paper, can
naturally be referred to as an $\alpha$-particle model, in contrast to
microscopic $\alpha$-cluster models where the intrinsic nucleonic
composition of the $\alpha$-particle is the basic ingredient 
\cite{che07,cuo13}.  This procedure follows very closely the one 
described and tested in \cite{gar14} for the two-body system of $^8$Be. 
In this way we can test the validity of the generalization to three-body systems.

The structure of $^{12}$C will be approximated as a three-alpha
system and calculated by use of the hyperspherical adiabatic expansion
method \cite{nie01}.  The continuum spectrum will be discretized by 
imposing a box boundary condition in the hyperradial coordinate, and 
the corresponding wave functions will be computed on the real
energy axis, without any particular treatment of the resonances.  All
states are equally treated as continuum states characterized by energy,
angular momentum, and parity.  A resonance would only be a
distribution of these discretized continuum states.

Two different alpha-alpha potentials will be used in our calculations, 
the Ali-Bodmer and the Buck potentials \cite{ali66,buc77}. 
These potentials are parametrized to
reproduce $\alpha-\alpha$ scattering properties and consequently the
two-body wave functions must be the same at large distances.  However,
both small distances and high partial wave properties can differ
drastically.  All the available versions of the Ali-Bodmer
potential reproduce the experimental $s$- and $d$-wave phase shifts, but only versions
'd' and 'e' reproduce  also the ones for $\ell=4$ \cite{ali66}. 
For this reason we have chosen to use the version ``d'' of the Ali-Bodmer
potential which has no bound states in contrast to the Buck potential \cite{buc77}.
Nevertheless, as shown in Ref.\cite{gar13}, Ali-Bodmer and Buck
potentials give rise to very similar phase shifts for $\ell=$~0,2,4,6, and 8.

The main difference between the two potentials is that the Buck potential contains 
two $s$-wave and one $d$-wave 
Pauli-forbidden bound states. This fact implies a very different short-distance 
behaviour for the $0^+$ and $2^+$ wave functions in $^8$Be, since the corresponding
radial wave functions show a different number of nodes depending on the potential used.  
For this reason, observable quantities sensitive to the short-distance structure,
like the photon emission, were expected to provide
information about the underlying two-body potentials. As shown in 
Ref.\cite{gar12}, to our surprise, we found no significant differences in the computed
bremsstrahlung cross sections for $^8$Be between these potentials.

At the three-body level things are very different. For instance, the ground state
in $^{12}$C is always the lowest computed 0$^+$ state, no matter which of the two potentials
is used. This is because when using the Buck potential we have two options.
Either we construct the phase equivalent alpha-alpha potentials \cite{gar99} or we 
exclude the Pauli forbidden adiabatic potentials before computing the radial three-body wave 
functions \cite{gar97}. In both cases the effective two-body potential shows a 
short-distance core repulsion, similar but not identical to the one in the Ali-Bodmer potential.
As a consequence, the structure of the three-body system looks very much the same
in all the cases. Therefore, we could off hand expect for $^{12}$C 
a dependence on the potential even smaller than in the $^8$Be case. This
point will be tested in our calculations.

The paper is organized as follows.
In Section II we describe the theoretical background needed to compute
the cross sections.  The particularization of the general expressions
to the three-body case is given in Section III, which is divided into
three subsections devoted to the three-body wave functions, the
transition matrix element, and the transition strengths,
respectively. In section IV we focus on the three-alpha system,
describing the details of the two-body potentials used
in the calculation and discussing the corresponding computed $^{12}$C
spectra. The electric quadrupole $\gamma$-emission cross sections are
described in Section V. Finally, in Section VI we give the computed
$E2$-transition strengths and compare with the predictions from the
rotational model. We finish the paper with the summary and the
conclusions.  Some derivations and expressions not essential for the
understanding of the paper, but important in order to make the paper
self-contained, have been collected in three appendices.

\section{Cross section expressions}

The photo-dissociation cross section for the breakup of a bound system $A$ with angular momentum $J_A$ into
a three-body system with angular momentum $J$ and three-body energy $E$ ($A+\gamma \rightarrow a+b+c$) 
is given by \cite{gar14,for03}:
\begin{equation}
\sigma_\gamma^{(\lambda)}(E)=\frac{(2\pi)^3 (\lambda+1)}{\lambda [(2\lambda+1)!!]^2}
\left(\frac{E_\gamma}{\hbar c} \right)^{2\lambda-1} \frac{d{\cal B}^{(\lambda)}}{dE}(J_A\rightarrow J),
\label{eq1}
\end{equation}
where $\lambda$ is the multipolarity of the electromagnetic transition, $E_\gamma$ is the photon energy
($E_\gamma=E+|B_A|$, where $B_A$ is the binding energy of $A$), and
\begin{equation}
\frac{d{\cal B}^{(\lambda)}}{dE}(J_A\rightarrow J)=
\frac{1}{2 J_A +1} \sum_i \left| 
\langle \Phi_J^{(i)}|| \hat{O}_\lambda || \Phi_{J_A} \rangle
\right|^2 \delta(E-E_i),
\end{equation}
where we have assumed that the continuum spectrum describing the final three-body system has been
discretized and the index $i$ runs over the discrete continuum states. The wave function of the
discrete state $i$, with energy $E_i$, is denoted by $\Phi^{(i)}_J$, and $\Phi_{J_A}$ is the wave 
function of the bound state $A$. The electromagnetic operator with multipolarity
$\lambda$ is denoted by $\hat{O}_\lambda$, which for the case of electric transitions reads:
\begin{equation}
\hat{\cal O}_{\lambda\mu}=e \sum_i Z_i r_i^\lambda Y_{\lambda \mu}(\Omega_i),
\label{oper}
\end{equation}
where $i$ runs over the charged particles in the system, each of them with charge $eZ_i$, and
$\bm{r}_i$ is the center of mass coordinate of particle $i$, whose direction is given by the angles
$\Omega_i$.

As described in appendix \ref{apen2}, the photo-dissociation cross section 
$\sigma_\gamma^{(\lambda)}(E)$ corresponding to the $A+\gamma \rightarrow a+b+c$ process,  
and the one corresponding to the inverse reaction $a+b+c \rightarrow A+\gamma$ 
(denoted now as  $\sigma^{(\lambda)}(E)$), are related through the expression:
\begin{equation}
\frac{\sigma^{(\lambda)}(E)}{\sigma_\gamma^{(\lambda)}(E_\gamma)}=
\nu! \frac{2 (2J_A+1)}{(2J_a+1)(2J_b+1)(2J_c+1)} 
\frac{32 \pi}{\kappa^5} \left(\frac{E_\gamma}{\hbar c} \right)^2 ,
\label{eq3}
\end{equation}
where $\nu$ is the number of identical particles in the three-body system, $J_a$, $J_b$, and $J_c$
are the total angular momenta of particles $a$, $b$, and $c$, respectively, and $\kappa$ is the three-body
momentum, which is defined as $\kappa=\sqrt{2mE/\hbar^2}$. The mass $m$ is the normalization 
mass used to define the Jacobi coordinates,  which are the coordinates usually employed to describe 
the three-body system \cite{nie01}.

As seen in Eq.(\ref{eq3}), the cross section $\sigma^{(\lambda)}(E)$ corresponding to the
radiative capture reaction $a+b+c \rightarrow A+\gamma$ has dimensions of length to the fifth power,
which corresponds to a surface in the six-dimensional space required to describe the incoming
three-body system. This cross section depends through
$\kappa$ on the normalization mass $m$. This is related to the fact that 
when using the hyperspherical coordinates (which are constructed from the Jacobi coordinates), the radial 
coordinate, the hyperradius, depends also on $m$. Therefore, a given value of the hyperradius will 
correspond to different three-body geometries (different relative distances between the three particles)
for different choices of $m$. As a consequence, the flux of incoming particles through a given hypersurface 
of radius $\rho$
will depend on $m$, and therefore also the cross section, which is the outgoing flux of particles normalized 
with the incoming flux (see appendix \ref{apen1}). Note that the well-defined physical observable, 
independent of the choice made for the normalization mass, is the reaction rate (see appendix \ref{apen2}).

If we now replace the bound state, $A$, by a continuum state with energy $E'$ and momentum $J'$, the 
photo-dissociation cross section for the process $A+\gamma \rightarrow a+b+c$ given in Eq.(\ref{eq1}) 
can be easily generalized as described in Ref.\cite{gar14}:
\begin{equation}
\frac{d\sigma_\gamma^{(\lambda)}}{dE'}(E)=
\frac{(2\pi)^3 (\lambda+1)}{\lambda [(2\lambda+1)!!]^2}
\left(\frac{E_\gamma}{\hbar c} \right)^{2\lambda-1} \frac{d{\cal B}^{(\lambda)}}{dEdE'}(J'\rightarrow J),
\label{eq4}
\end{equation}
where $E_\gamma=E-E'$ is the photon energy and 
\begin{eqnarray}
\lefteqn{
\frac{d{\cal B}^{(\lambda)}}{dEdE'}(J'\rightarrow J)= }  \label{eq5}\\ & &
\frac{1}{2 J' +1} \sum_{i,j} \left| 
\langle \Phi_J^{(i)}|| \hat{O}_\lambda || \Phi_{J'}^{(j)} \rangle
\right|^2 \delta(E-E_i) \delta(E'-E'_j),
\nonumber
\end{eqnarray}
where the initial continuum states have also been discretized, and the index $j$ runs over 
the initial continuum discrete states with energy $E'_j$ and wave function $\Phi_{J'}^{(j)}$.
It is important to note that the summation over $i$ and $j$ in the equation above is not 
unrestricted, but limited to the initial and final energy ranges of experimental interest.
We will come back to this point later on.

Finally, according to Eq.(\ref{eq3}), we have that the cross section for the continuum to 
continuum reaction $a+b+c \rightarrow A +\gamma$, with initial and final energies $E$ and $E'$, 
and initial and final angular momenta $J$ and $J'$ is given by:
\begin{eqnarray}
\lefteqn{
\frac{d\sigma^{(\lambda)}}{dE'}(E)= } \label{eq6} \\ & &
\nu! \frac{2 (2J_A+1)}{(2J_a+1)(2J_b+1)(2J_c+1)} 
\frac{32 \pi}{\kappa^5} \left(\frac{E_\gamma}{\hbar c} \right)^2 
\frac{d\sigma_\gamma^{(\lambda)}}{dE'}(E_\gamma).
\nonumber
\end{eqnarray}

As discussed in Ref.\cite{gar14}, the total bremsstrahlung cross
section, as a function of the incident energy $E$, is obtained after
integration over $E'$ (or over the photon energy $E_\gamma$). In our
description using discrete initial and final continuum states, the
integral over $E'$ is trivially made thanks to the $\delta(E'-E'_j)$
function in Eq.(\ref{eq5}), and just a summation over the index $j$
running over the final discrete states remains:

\begin{eqnarray}
\label{totcs}
\sigma^{(\lambda)}(E)= \frac{\nu!}{(2J_a+1)(2J_b+1)(2J_c+1)} 
\left(\frac{E_\gamma}{\hbar c} \right)^{2\lambda+1}  \\ \nonumber
 \times \frac{16}{\pi}\frac{1}{\kappa^5}
\frac{(2\pi)^5 (\lambda+1)}{\lambda [(2\lambda+1)!!]^2}
\sum_{i,j} \left| 
\langle \Phi_J^{(i)}|| \hat{O}_\lambda || \Phi_{J'}^{(j)} \rangle 
\right|^2 \delta(E-E_i) .
\end{eqnarray}

As anticipated below Eq.(\ref{eq5}) and stated in Ref.\cite{gar12}, 
the computed cross sections should be
obtained in close analogy to the experimental setup, where only a
finite range of final relative energies is measured, usually around a
resonance in the final system. This means that the integral over $E'$
has to be performed only over this precise energy range. In our
language of discrete continuum states, this means that the summation
over $j$ in the equation above runs only over the final discrete
states whose energy $E'_j$ is contained in the chosen final energy
window. It is obvious then that the total bremsstrahlung cross section
depends on such final energy window, see \cite{gar13,gar14}.  In other
words, definitive statements about resonance properties must take into
account that these continuum states have an energy at best defined
with an accuracy of less than its width.

\section{Three-body ingredients}

In order to compute the cross section given in Eq.(\ref{totcs})
we first need the wave functions for the three interacting particles
in states with given energy and specified angular momentum and parity.
Then we need the transition probability from one state to
another. This requires specific matrix elements which by combination
of various kinematic factors provide cross sections and through
proper definitions also the related transition strengths.

\subsection{Three-body wave functions}
\label{3bdwf}

In this work we shall construct the three-body wave functions using the hyperspherical adiabatic
expansion method described in Ref.\cite{nie01}. In this method the wave function is expanded in terms of a
complete set of angular functions $\{\phi_n^J\}$, where $J$ is the total angular momentum of the three-body
system:
\begin{equation}
\Phi_J={1\over\rho^{5/2}} \sum_n f^J_n(\rho) \phi_n^J(\rho,\Omega);
(\Omega\equiv\{\alpha, \Omega_{x}, \Omega_{y} \}), 
\label{eq7}
\end{equation}
where $\rho=\sqrt{x^2+y^2}$, $\alpha=\arctan({x/y})$, and $\{ \Omega_{x},\Omega_{y}\}$ are the angles 
defining the
directions of $\bm{x}$ and $\bm{y}$, which are the Jacobi coordinates used to describe the
system.  Writing the Schr\"odinger equation in terms of these coordinates, they can be separated into
angular and radial parts:
\begin{equation}
\hat{\Lambda}^2 \phi_n^J+\frac{2 m \rho^2}{\hbar^2} 
\left(V_{jk}+V_{ik}+V_{ij}  \right) \phi_n^J   =
\lambda_n(\rho) \phi_n^J,
 \label{eq8} 
\end{equation}
and
\begin{eqnarray}
\left[ -\frac{d^2}{d\rho^2} +  \frac{2m}{\hbar^2} (V_{3b}(\rho) - E)
+ \frac{1}{\rho^2}
\left( \lambda_n(\rho)+\frac{15}{4} \right) \right] f^J_n(\rho) \nonumber &&\\
& \hspace*{-8cm}
+ \sum_{n'} \left( -2 P_{n n'} \frac{d}{d\rho} - Q_{n n'} \right)f_{n'}(\rho)
= 0, &  \label{eq9}
\end{eqnarray}
where $V_{jk}$, $V_{ik}$, and $V_{ij}$ are the two-body interactions between each pair of particles, 
$\hat{\Lambda}^2$ is the hyperangular operator, see Ref.\cite{nie01}, and
$m$ is the normalization mass. In Eq.(\ref{eq9}) $E$ is the three-body
energy, and the coupling functions $P_{n n'}$ and $Q_{n n'}$ are given
for instance in \cite{nie01}. The potential $V_{3b}(\rho)$ is used for
fine tuning to take into account all those effects that go beyond the
two-body interactions.

It is important to note that the angular functions used in the
expansion in Eq.(\ref{eq7}) are precisely the eigenfunctions of the angular
part of the Schr\"odinger (or Faddeev) equation(s). Each of them is in practice obtained
by expansion in terms of the hyperspherical harmonics, see Eq.(\ref{ap32}). Obviously this
infinite expansion has to be truncated at some point, maintaining only
the contributing components.

The eigenvalues $\lambda_n(\rho)$ in Eq.(\ref{eq8}) enter in the
radial equations (\ref{eq9}) as a basic ingredient in the effective
radial potentials. Accurate calculation of the $\lambda$-eigenvalues
requires, for each particular component, a sufficiently large number
of hyperspherical harmonics. In other words, the maximum value of the
hypermomentum, $K_{max}$, for each component must be large enough to
assure convergence of the $\lambda$-functions in the region of
$\rho$-values where the $f^J_n(\rho)$ wave functions are relevant for
the calculation of the electromagnetic operator matrix element.

Finally, the last convergence to take into account is the one
corresponding to the expansion in Eq.(\ref{eq7}). Typically, for bound
states, this expansion converges rather fast, and usually three or
four adiabatic terms are already sufficient.

In our calculations the continuum is discretized by use of a box
boundary condition.  This means that the radial wave functions
$f_n(\rho)$ are imposed to be zero at a given maximum value of $\rho$,
which is typically taken equal to a few hundreds of fm
($\rho_{max}=200$ fm in our calculations). No distinction is made
between resonances and ordinary continuum states. Therefore, in order
to place the initial and final energy windows matching the resonance
energies, it will be necessary to have some previous information about
what these energies are.  It also implies that the computed result
simultaneously includes on-resonance as well as continuum background
contributions.

\subsection{Transition matrix element}

Once the initial and final three-body wave functions are computed, we can now obtain the
square of the transition matrix element
\begin{equation}
|\langle \Phi_J^{(i)} || \hat{\cal O}_\lambda || \Phi_{J'}^{(j)} \rangle|^2=
(2J+1)\sum_{\mu M'}
|\langle \Phi_{JM}^{(i)} | \hat{\cal O}_{\lambda \mu} | \Phi_{J'M'}^{(j)} \rangle|^2
\label{eq12}
\end{equation}
which enters in Eq.(\ref{totcs}), and where $\hat{\cal O}_{\lambda \mu}$ is the electric multipole operator
given in Eq.(\ref{oper}).

From the definition of the Jacobi coordinates, Ref.\cite{nie01}, it is not difficult to see that the vector
$\bm{r}_p$ giving the position of particle $p$ can be written as:
\begin{equation}
\bm{r}_p=\sqrt{\frac{m(m_q+m_s)}{m_p(m_p+m_q+m_s)}} \bm{y}_p
\end{equation}
where $m_p$, $m_q$, and $m_s$ are the masses of the three particles and $\bm{y}_p$ is the Jacobi
coordinate defined between particle $p$ and the center of mass of the other two.
Therefore, using Eq.(\ref{oper}), we get:
\begin{eqnarray}
\langle \Phi_J^{(i)} || \hat{\cal O}_\lambda || \Phi_{J'}^{(j)} \rangle & = &
e\sum_{p=1}^3 Z_p 
\left( \frac{m(m_q+m_s)}{m_p(m_p+m_q+m_s)}    \right)^{\lambda/2} \nonumber \\ & & \times
\langle \Phi_J^{(i)} || y_p^\lambda Y_{\lambda}(\hat{r}_p) || \Phi_{J'}^{(j)} \rangle,
\end{eqnarray}
and the calculation of the reduced matrix element in Eq.(\ref{eq12})
requires only the calculation of the matrix element $\langle
\Phi_J^{(i)} || y_p^\lambda Y_{\lambda}(\hat{r}_p) || \Phi_{J'}^{(j)}
\rangle$ for each of the three possible definitions of the Jacobi
coordinate $\bm{y}_p$. This reduced matrix element is obtained as described 
in appendix \ref{apen3}. The final expression in Eq.(\ref{ap34}) 
is rather elaborate although mostly due to the Racah
algebra necessary to account correctly for all total angular momentum
couplings in general cases of particles with finite intrinsic spin.

The integral over $\rho$ in Eq.(\ref{ap34}) involves the radial wave
functions $f_n^J$ and $f_{n'}^{J'}$ corresponding to the initial and
final continuum states. Therefore the function to be integrated does
not fall to zero at infinity, but instead gives rise to an apparent divergence
of the necessary integrals.  This divergence is obviously nonphysical
and mathematically ill-defined until a suitable limiting procedure is
chosen.  A simple way to solve this numerical problem was proposed in
Ref.\cite{gar14}, where the integrand was multiplied by the factor
$e^{-\eta^2 \rho^2}$, in such a way that the correct result is obtained in
the limit of $\eta=0$.  In practice, a value of $\eta$ in the vicinity
of $\eta=0.01$ fm$^{-1}$ is enough to get a sufficient accuracy.

It is important to note that the divergence mentioned in the previous
paragraph appears due to the application of the long-wavelength approximation,
thanks to which the electric field can be obtained from the charge 
density only (Siegert theorem). This theorem leads to the
electric transition multipole operator given in Eq.(\ref{oper}).
In Ref.\cite{doh13} an extension of the Siegert theorem not relying 
on the long-wavelength approximation was proposed,
in such a way that the divergence problem in the matrix element 
disappears. The equivalence between this procedure and 
different techniques designed to treat the divergence problem
(the one used in this work among them) has been investigated
in Ref.\cite{doh14}.

\subsection{Transition strength}
\label{trans}

In principle, calculation of the ${\cal B}^{(E\lambda)}$ strength for the
continuum to continuum transition, $a+b+c \rightarrow A +\gamma$, could be
made directly through Eq.(\ref{eq5}), which thanks to the $\delta$-functions
would permit to obtain the total transition strength just by summation 
over $i$ and $j$ of the square of the reduced matrix element. This would
in fact be the same procedure as when a bound state is involved in the
transition. However, as discussed in Ref.\cite{gar13}, an indiscriminate sum over 
initial and final continuum states makes the result rather meaningless. The 
information about resonance properties
is completely washed out and, even worse, weighted at the wrong
energies.  Furthermore, due to the undesired divergence produced by the
soft-photon contribution, which appears when the energy of the emitted
photon approaches zero ($E^\prime \rightarrow E$ or
$E^\prime_j \rightarrow E_i$) \cite{gar12}, the calculation itself is pretty
complicated. For these reasons, we shall obtain the transition strength as described 
in Ref.\cite{gar13}, i.e. directly from the cross section in Eqs.(\ref{eq6}) and (\ref{eq4}).
Two different methods will be used. 

In the first method the strength is obtained from the total (integrated) cross section. 
More precisely, for a transition
between some initial and final energy windows, typically around some resonances
in the initial and final states, integration of Eq.(\ref{eq6}) over
$E$ and $E'$ within those two windows will provide the total cross section
for the transition. If the photon energy, $E_\gamma$, were constant, this would
immediately provide the total transition strength just after division by the constants
multiplying the transition strength. However, since this is not correct, 
we must use an average value for $E_\gamma^{2\lambda+1}$. In particular, the
photon energy will be taken as $E_\gamma=E_c-E'_c$ where $E_c$ is the energy of the
cross section peak corresponding to the resonance in the initial state and
$E'_c$ is the center of the final energy window (usually the resonance energy in the
final state).
Of course, this procedure assumes information about resonance positions, and it can
be sensitive to rather small variations around a chosen $E_\gamma$ owing to the power of $(2\lambda+1)$
for ${\cal B}(E\lambda)$ transitions, see Ref.\cite{gar12} for details.

The second method exploits the fact that in the vicinity of a resonance the cross section of the 
photo-dissociation reaction $A+\gamma \rightarrow a+b+c$ takes the form
\begin{equation}
\sigma_\gamma(E_\gamma)=\frac{2J+1}{2(2J_A+1)} \frac{\pi \hbar^2 c^2}{E_\gamma^2}
\frac{\Gamma_R \Gamma_\gamma}{(E-E_R)^2+\Gamma^2/4},
\label{bw1}
\end{equation}
where after the collision the particles $a$, $b$, and $c$ are assumed to populate a resonance 
with angular momentum $J$, energy $E_R$, and width for decay into three particles $\Gamma_{R}$. 
$J_A$ is the angular momentum of $A$ and $\Gamma=\Gamma_R+\Gamma_\gamma$, where $\Gamma_\gamma$
is the $\gamma$-decay width of the three-body resonance.

Using Eq.(\ref{eq3}), we can easily obtain the expression equivalent to Eq.(\ref{bw1})
for the inverse cross section after the three-body collision $a+b+c \rightarrow A+\gamma$, which 
can be written as
\begin{eqnarray}
\lefteqn{ \sigma(E)= }  \label{bw2} \\ & &
 \frac{\nu!(2J+1)}{(2J_a+1)(2J_b+1)(2J_c+1)} \frac{8(2\pi)^2}{\kappa^5}
\frac{\Gamma_R \Gamma_\gamma}{(E-E_R)^2+\Gamma^2/4},
\nonumber
\end{eqnarray}
where now the three colliding particles, with spins $J_a$, $J_b$ and $J_c$ are assumed to populate
a three-body resonance 
with angular momentum $J$, energy $E_R$, and width for decay into three particles $\Gamma_{R}$.

The second method uses the value of $\Gamma_\gamma$ in the equation above in order to fit the peak 
in the computed cross section in Eq.(\ref{totcs}) corresponding to the three-body resonance at $E=E_R$,
 in such a way that from $\Gamma_\gamma$ we can obtain the transition strength thanks to the
well-known expression \cite{sie87}:
\begin{equation}
\Gamma_\gamma=\frac{8\pi (\lambda+1)}{\lambda[(2\lambda+1)!!]^2}
\left( \frac{E_\gamma}{\hbar c} \right)^{2\lambda+1} {\cal B}^{(E\lambda)}(J\rightarrow J_A).
\label{gamma}
\end{equation}

The two methods can be compared. The first prescription depends on the
windows chosen for both initial and final states.  These choices
should be made precisely to reproduce the conditions in a given
measurement.  However, then the average photon energy becomes
important.  The second prescription relies on the behavior of the
cross section for energies around the resonance peak.  This dependence
is assumed to have the Breit-Wigner shape with the advantage (and
related weakness) that only a few points around the peak are used.
The photon energy does not enter as a multiplicative factor, but
deviations from the assumed simple Lorentzian behavior are not
accounted for.

\section{The three-alpha system}

In this section we describe the calculation made to construct the wave functions for a system of
three alpha particles. We start giving the details of the two-body $\alpha$-$\alpha$ potentials used  
in the calculation and some properties of the $^8$Be-spectrum. In the second part we summarize the
properties of the $^{12}$C-spectrum.

\subsection{$^8$Be properties}
\label{pots}

We shall consider the same two $\alpha$-$\alpha$ potentials used in Ref.\cite{gar13},
i.e. the Buck potential \cite{buc77} and version $d$ of the Ali-Bodmer potential given in 
Ref.\cite{ali66}. The Buck potential has two spurious deep-lying $\alpha$-$\alpha$ bound states 
for $s$-waves and one more for $d$-waves. These spurious states correspond to Pauli forbidden states.
On the contrary, the Ali-Bodmer potential is a shallow potential not holding any bound $\alpha$-$\alpha$
state. These two potentials give rise to very similar $\ell=$0, 2, 4, 6, and 8 phase shifts.

\begin{table*}
\begin{center}
\caption{Properties of the five lowest computed resonances in
  $^8$Be. The first two rows give, when available, the corresponding
  experimental energies, $E_r$, and widths, $\Gamma_r$, taken from
  Ref.\cite{til04}. The computed values with the Buck and Ali-Bodmer
  potentials are given by the third and fourth rows, and by the fifth
  and sixth rows, respectively.  All the energies and widths are given
  in MeV. The following four rows give, also for the two
  $\alpha-\alpha$ potentials, the real and imaginary parts of
  $\sqrt{<r^2>}$, computed with the complex scaling method. These
  values are given in fm.  }
\label{tab1}
\begin{ruledtabular}
\begin{tabular}{|c|ccccc|}
\hline
 $J^+$  &  $0^+$ & $2^+$ & $4^+$ &  $6^+$ & $8^+$ \\
\hline
$E_r$ (Exp.)        &         0.0918          & $2.94\pm0.01$   &   $11.35\pm0.15$      &  --   &  --      \\
$\Gamma_r$  (Exp.)  & $(5.57\pm0.25) 10^{-6}$ & $1.51\pm0.02$   &     $\sim 3.5  $      &  --   &  --      \\
\hline
 $E_r$ (Buck)       &          0.091           &      2.88       &      11.78            &  33.55  &  51.56   \\
 $\Gamma_r$ (Buck)  &   $3.6\cdot10^{-5}$      &      1.24       &       3.57            &  37.38  &  92.38   \\
\hline
 $E_r$  (Ali-Bodmer d)   &          0.092           &      2.90       &      11.70            &  34.38  &  53.65   \\
$\Gamma_r$ (Ali-Bodmer d)&   $3.1\cdot10^{-6}$      &      1.27       &       3.07            &  37.19  &  93.74   \\
\hline
Re{$\sqrt{<r^2>}$} (Buck) & 5.61 & 3.51 & 2.93 & 2.82 & 2.76 \\
Im{$\sqrt{<r^2>}$} (Buck) & 0.01 & 1.29 & 0.82 & 1.44 & 1.77 \\
\hline
Re{$\sqrt{<r^2>}$} (Ali-Bodmer d) & 5.80  & 3.58 & 2.91 & 2.70 & 2.73 \\
Im{$\sqrt{<r^2>}$} (Ali-Bodmer d) & 0.001 & 1.24 & 0.76 & 1.40 & 1.73 \\
\hline
\end{tabular}
\end{ruledtabular}
\end{center}
\end{table*}

The spectrum of $^8$Be obtained with these two potentials is discussed in Ref.\cite{gar13}. In here
we only summarize in Table~\ref{tab1} the energies and widths of the different states. 
The first two rows show the known experimental energies and widths \cite{til04}
of the $0^+$, $2^+$, and $4^+$ resonances in $^8$Be. The experimental values are well reproduced
by both potentials. The computed resonances have been obtained
as poles of the ${\cal S}$-matrix by use of the complex scaling method \cite{ho83,moi98}.
The widths of the computed $6^+$ and $8^+$ resonances are very big, comparable to their
energies, and actually, they should not be considered as well-defined resonances.

After a complex scaling calculation the complex rotated wave functions
of the resonances fall off asymptotically as ordinary bound states. It
is therefore possible to compute mean-square-radii, expectation values
of $r^2$, which for resonances are complex numbers, in contrast to the
real values obtained for bound states even if the corresponding wave
functions have been complex rotated.  The real and imaginary parts of
$\langle r^2 \rangle^{1/2}$ for each of the computed resonances are
also given in Table~\ref{tab1}. As introduced by T. Berggren \cite{ber68,ber96}, 
and also discussed in Ref.\cite{moi98}, the
real part of the expectation value of a given complex rotated operator
has been attempted interpreted as a corresponding average value over
continuum wave functions in a range of energies around the
resonance. It is then tempting to associate the imaginary part of
$\langle r^2 \rangle^{1/2}$ with an uncertainty of the resonance size
arising from the non-zero width of the state.

\subsection{$^{12}$C spectrum}

The resonances of the three-alpha system are obtained as described in
Ref.\cite{alv07}. The method follows the hyperspherical adiabatic
expansion method \cite{nie01} sketched in Sect.~\ref{3bdwf}, which is used
in combination with the complex scaling method \cite{ho83,moi98}. The
three-body resonances appear then as ordinary bound states with
complex energy, whose real and imaginary parts give the resonance
energy and half the width of the resonance. This method does not make
any assumption about the resonance properties. For instance,
the resonance decay mechanism is dictated by the dynamic evolution of the resonances, 
i.e., by the change in structure from small to large distances. In this way, 
the sequential and direct decay channels are both simultaneously taken 
into account, and the corresponding branching ratios are directly dictated 
by the resonance wave function \cite{alv08}.

In order to reproduce the known experimental energies in the
$^{12}$C-spectrum a fine tuning with a short-range three-body force is
required. This is done by the potential $V_{3b}(\rho)$ introduced in
the set of radial equations given in Eq.(\ref{eq9}). As in
Ref.\cite{alv07}, we shall consider here a gaussian three-body
potential $V_{3b}=Se^{-\rho^2/b^2}$, where the range $b$ is taken
equal to 6 fm, which approximately corresponds to the hyperradius
obtained from three touching $\alpha$-particles. This $V_{3b}$
construction maintains the structure of a three-body state, but
varying the strength, $S$, the energy position can be adjusted to
reproduce the measured value of the resonance.

When using the Buck potential, due to the existence of Pauli forbidden two-body states, a direct
three-body calculation gives rise to a large amount of spurious bound three-body
states. To avoid this problem, the three-body calculation is made using the phase-equivalent
version of the Buck potential. This potential is constructed numerically from the original one, and it
provides a two-body potential with exactly the same phase shifts for all energies, but where the
undesired forbidden bound states have been removed from the two-body spectrum, see Ref.\cite{gar99}
for details. 

\begin{table*}
\begin{center}
\caption{Calculated and measured energies $E_R$ (in MeV) and partial $\alpha$-decay widths
$\Gamma_R$ (in keV) of the $^{12}$C resonances for different $J^\pi$. Experimental values
(labeled ``exp'') are from \cite{ajz90,dig06,zim13,zim13b,fre11}. The labels ``AB'' and ``Buck'' refer to the
calculations obtained using the Ali-Bodmer and Buck $\alpha$-$\alpha$ potentials specified
in Sect.~\ref{pots}. The strength $S$ (in MeV) used in the gaussian three-body potential 
$V_{3b}$ is also given for each of the calculations (the strength of the three-body 
force is always taken to be $b=6.0$ fm). The energies are measured from the three-alpha threshold.} 
\label{tab2}
\begin{ruledtabular}
\begin{tabular}{|c|cc|ccc|ccc|}
  $J^\pi$ &  $E_{R,\mbox{\scriptsize exp}}$   &   $\Gamma_{R,\mbox{\scriptsize exp}}$   &
   $E_{R,\mbox{\scriptsize AB}}$   &   $\Gamma_{R,\mbox{\scriptsize AB}}$   &   $S$   &  
   $E_{R,\mbox{\scriptsize Buck}}$ &   $\Gamma_{R,\mbox{\scriptsize Buck}}$ &   $S$     
                                         \\ \hline
 $0^+_1$ & $-7.275^{(a)}$          & --                    & $-7.27$   &  --             & $-22.6$ & $-7.27$ &    --       & $-22.0$ \\
 $0^+_2$ & $0.380^{(a)}$           & $0.009\pm0.001^{(a)}$ &    0.38         & $\lesssim 0.05$ &      $-18.2$    &      0.38       & $\lesssim 0.05$ &     $-18.0$     \\
 $0^+_3$ & $4.20\pm0.14^{(b)}$     & $3440\pm220^{(b)}$    &    5.52        &      2200        &      $-18.2$    &      4.12       &      700    &     $-18.0$     \\   \hline
 $2^+_1$ & $-2.8356\pm0.0003^{(a)}$&     0.0               & $-2.84$        &     --          & $-12.6$          & $-2.85$   &      --     &  $-12.1$  \\
 $2^+_2$ &   $2.76\pm0.11^{(c)}$   & $800\pm130^{(c)}$     &   1.75/2.36     &    375/1150     & $-12.6$/$-3.0$  &  1.72/2.35      &    161/920  &  $-12.1$/$-3.0$  \\
         &   $2.86\pm0.05^{(d)}$   & $2080\pm300^{(d)}$    &                 &                 &                 &                 &             &                  \\
 $2^+_3$ &   $3.88\pm0.05^{(a)}$   & $430\pm80^{(a)}$      &   3.90/6.87     &    163/2010       & $-12.6$/$-3.0$  &  5.04/5.48      &   570/3600 &  $-12.1$/$-3.0$  \\  \hline
 $4^+_1$ &   $6.0\pm0.2^{(e)}$   & $1700\pm200^{(e)}$      &      5.33        &    3700       &      16.3       &      5.48       &     3800     &      13.4        \\   
 $4^+_2$ &   $6.81\pm0.02^{(a)}$   & $258\pm15^{(a)}$      &      6.82        &    620        &      16.3       &      6.81       &     800      &      13.4        \\  
 $4^+_3$ &                         &                       &      13.1        &   1800        &      16.3       &      13.1       &     2060     &      13.4        \\  
\end{tabular}
$(a)$ Ref.\cite{ajz90}; $(b)$ Ref.\cite{dig06}; $(c)$ Ref.\cite{zim13}; $(d)$ Ref.\cite{zim13b}, $(e)$ Ref.\cite{fre11}
\end{ruledtabular}
\end{center}
\end{table*}

In Table~\ref{tab2} we give the computed energies and widths obtained for the $0^+$, $2^+$,
and $4^+$ states, which are the states of interest for this work.
The experimental data are taken from Refs.\cite{ajz90,dig06,zim13,zim13b,fre11}. 
The results obtained with the Ali-Bodmer and the Buck potentials are given.
For each calculation the value of the strength, $S$, used in the three-body force is also given.

For the $0^+$ states, if the three-body force is used to fit the energy of the ground state, we then
get a second $0^+$ state (the Hoyle state) slightly bound, which is clearly incorrect. In order to
get the correct energy for the Hoyle state, whose structure is expected to agree more with the 
three-alpha model than the ground state, it is therefore necessary to weaken the three-body 
attraction, which makes the ground state underbound by a bit more than 1 MeV. When this
is done, a third $0^+$ state appears at an energy of about 5.5 MeV with the Ali-Bodmer potential and 4 MeV
with the Buck potential. This last energy agrees with the experimental value, although the large 
experimental width is better reproduced with the Ali-Bodmer potential. It is important to keep in mind 
that the resonances are computed as poles of the ${\cal S}$-matrix, and the experimental data are 
often obtained after an ${\cal R}$-matrix analysis of the cross sections. These two different procedures 
can lead to sometimes very different values for the resonance widths \cite{bar08}.
In Table~\ref{tab2} the two values of the strength of the three-body force $S$ fitting the experimental 
energies of the $0^+_1$ and $0^+_2$ states, are given.  

For the $2^+$ states we proceed in a similar way. When the three-body
force is used to fit the energy of the bound $2^+$ state, we then get
the second $2^+$ state with an energy and width of $(1.75,0.38)$ MeV
and $(1.72,0.16)$ MeV with the Ali-Bodmer and Buck potentials,
respectively.  These values of energy and width are lower than the
experimental values recently given in \cite{zim13,zim13b}.  Again, in
order to fit this energy we have to weaken the three-body force, in
such a way that the $2^+_2$ resonance appears at about 2.4 MeV. An
even weaker three-body force would place the $2^+_2$ state at 2.8 MeV,
in better agreement with the experimental value, but in this case the
resonance would be too wide and more difficult to obtain through a
complex scaling calculation (due to the large rotation angle required
in this case).  When the energy of the $2^+_2$ resonance is placed at
about 2.4 MeV, the corresponding width is around 1 MeV with both, the
Ali-Bodmer and the Buck potential. This width agrees with the value
given in Ref.\cite{zim13}, but it is a factor of 2 smaller than the
value given in Ref.\cite{zim13b}. Note that the resonance widths given
in Refs.\cite{zim13} and \cite{zim13b} are obtained from the same
experiment, although the fit of the data is obviously different.  This
fact emphasizes that not too much confidence should be placed on the
comparison in Table~\ref{tab2} between the computed and experimental
widths.  

Beside the two lowest $2^+$ (bound and resonance) states, also a
$2^+_3$ state appears in the calculations.  As seen in Table~\ref{tab2},  
when the energy of the $2^+_2$ resonance is moved from $\sim 1.8$ MeV to $\sim 2.4$ 
MeV, the effect on the $2^+_3$ state depends rather strongly on the potential
used. With the Ali-Bodmer potential the energy moves from 3.90 MeV
to 6.87 MeV, and the width changes quite dramatically from a rather small
value, 0.16 MeV, to 2.01 MeV. However, with the Buck potential 
basically only the resonance width changes, from 0.6 MeV to 3.6 MeV, while
the energy value remains quite stable, since it only changes from 5.04 MeV 
to 5.48 MeV. 

Concerning the 4$^+$ states, the existence of a resonance at 6.81 MeV above
the three-body threshold, with a width of 0.258 MeV has been known for a long 
time (Ref.\cite{ajz90}). Much more recently, an additional $4^+$ state has been 
reported with an energy of 6.0 MeV (above threshold) and a width of 1.7 MeV \cite{fre11}. Our 
calculations are consistent with the existence of a well-defined and relatively narrow
resonance, which by means of the three-body force can be made to appear at 6.81
MeV, in agreement with the experimental value for the $4^+_2$ state. The computed
width for this resonance is 0.62 MeV with the Ali-Bodmer potential and 0.80 MeV with the
Buck potential. When this is done it can be seen that a quite broad resonance 
($\sim 3.7$ MeV wide) appears numerically with both potentials at and energy of
about $\sim 5.4$ MeV. This broad resonance can be interpreted as the one 
given experimentally in Ref.\cite{fre11}. Also, both the Ali-Bodmer and Buck
potentials, give rise to a $4^+_3$ state at 13.1 MeV, having also a similar width
in the vicinity of 2.0 MeV.

As seen in Table \ref{tab2}, the energies and widths of the resonances are rather independent
of the choice made for the potential. In general, with the two potentials used, a similar three-body
force gives rise to pretty much the same resonance properties. However, there are a few
exceptions, mainly the $0^+_3$ and $2^+_3$ states, and the narrow $2^+_2$ resonance at
1.7 MeV, which with the Ali-Bodmer potential is more than a factor of two broader than with
the Buck potential. These differences can lead to visible differences in the differential
cross sections, and therefore also in the transition strengths, for the cases where these 
potential-dependent resonances are involved.

\begin{figure}
\epsfig{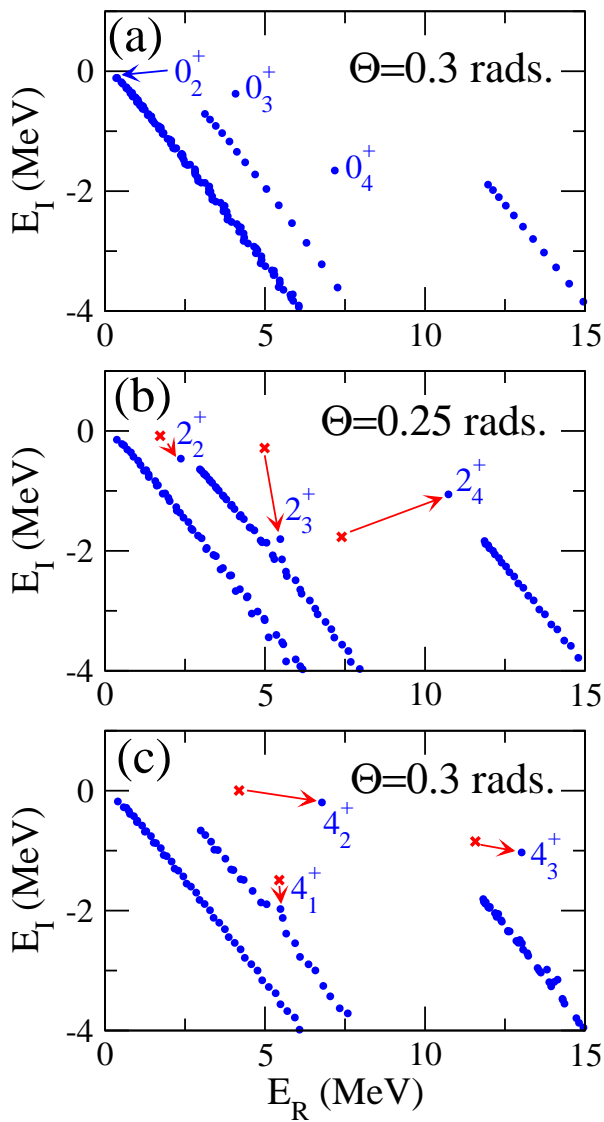}
\caption{(Color online)
Fourth quadrant of the energy plane showing the discrete $0^+$, $2^+$, and $4^+$ continuum 
spectra in $^{12}$C after a complex scaling calculation with the Buck potential. The angle $\theta$
on each panel is the scaling angle used in the corresponding calculation. The strength of 
the three-body force is $-18.0$ MeV, $-3.0$ MeV, and $13.4$ MeV for the $0^+$, $2^+$, and $4^+$ 
states, respectively.  The bound $0^+_1$ and $2^+_1$ states are not shown. The lowest
resonances for each of the three angular momenta are indicated by the corresponding labels. 
In the $2^+$ and $4^+$ cases the red crosses indicate the position of the resonances when the 
strength of the three-body force is reduced from $-3.0$ MeV to $-12.1$ MeV, and from 13.4 MeV 
to 7 MeV, respectively, and the arrows show how the resonances move when decreasing the three-body
attraction.}
\label{fig1}
\end{figure}

\begin{table*}
\begin{center}
\caption{ Computed values of $\rho_{rms}=\langle \rho^2 \rangle^{1/2}$ (in fm) and $r_{rms}$ (in fm) for the same 
$^{12}$C states shown in Table~\ref{tab2}.} 
\label{tab3}
\begin{ruledtabular}
\begin{tabular}{|c|cc|cc|}
  $J^\pi$ & $\rho_{rms}$ (AB) &  $r_{rms} (AB)$  & $\rho_{rms}$ (Buck)  &  $r_{rms}$ (Buck)
                                         \\ \hline
 $0^+_1$ & 6.9                  & 2.5                 &     6.8   & 2.5     \\
 $0^+_2$ & $11.5$             & $3.6$              & $11.4$  & $3.6$  \\
 $0^+_3$ & $11.3+i4.0$              & $3.5+i1.1$              & $10.9+i2.3$ & $3.4+i0.6$  \\   \hline
 $2^+_1$ &  6.8                &   2.5               &      6.7  &   2.4        \\
 $2^+_2$ & $9.8+i3.9$/$10.7+i2.4$  & $3.2+i1.0$/$3.6+i0.7$ & $9.9+i2.9$/$10.0+i2.6$ & $3.2+i0.7$/$3.2+i0.7$  \\
 $2^+_3$ & $6.0+i3.1$/$6.9+i0.9$   & $2.2+i0.7$/$2.5+i0.2$ &  $8.3+i1.0$/$7.5+i6.8$ & $2.8+i0.2$/$2.4+i1.7$   \\   \hline
 $4^+_1$ & $9.3+i1.7$    & $3.1+i0.4$  & $9.5+i1.6$    & $3.1+i0.4$  \\ 
 $4^+_2$ & $6.0+i0.1$    & $2.3+i0.1$  & $5.9+i0.1$   & $2.3+i0.1$  \\   
 $4^+_3$ & $7.7+i1.6$    & $2.7+i0.4$  & $7.9+i1.5$   & $2.7+i0.4$  \\   
\end{tabular}
\end{ruledtabular}
\end{center}
\end{table*}

As an illustration of how the resonance properties have been obtained, we show now in 
Fig.~\ref{fig1} the discretized continuum spectra after a
complex scaling calculation for the $0^+$ (Fig.~\ref{fig1}a), $2^+$ (Fig.~\ref{fig1}b), and
$4^+$ (Fig.~\ref{fig1}c) states in $^{12}$C. We show only the results obtained with the Buck potential.
The calculations have been done using a complex scaling angle, $\theta$, of 0.30 rads for the 
$0^+$ and $4^+$ states, and 0.25 rads for the $2^+$ states. In all the three cases the energy 
cuts starting from the origin, from the $2^+$ resonance in $^8$Be, and from the $4^+$ resonance 
in $^8$Be are clearly seen. 
Note that, due to the very small energy of the $0^+$ resonance in $^8$Be, the energy cut starting
from the $0^+$ state overlaps with the cut starting from the origin, which corresponds
to strict three-body continuum states.

These cuts are rotated in the complex energy plane by an angle
$2\theta$, and the states in the cuts correspond to pure continuum states. Typically, the states
corresponding to the three-body resonances fall clearly out of the energy cuts, and their position
is independent of the scaling angle $\theta$. These resonances are indicated in the figure  
with the corresponding labels
(the bound $0^+_1$ and $2^+_1$ states are not shown in the figure). However, there are 
several cases where the resonances lie pretty close to the continuum
states, and their identification is not so obvious. The natural way to isolate these resonances
from the continuum states would be to increase the value of the scaling angle $\theta$. However this
is not always an efficient solution, due to the technical difficulties arising from the
use of a too large scaling angle. For narrow resonances close to the threshold, like the Hoyle state
($0^+_2$ state), the resonance wave function falls exponentially to zero very fast, and the
resonance can easily be identified by looking directly into the resonance wave function. For
not very narrow resonances, like the $2^+_3$ state in Fig.~\ref{fig1}b, or the $4^+_1$ in 
Fig.~\ref{fig1}c, it is much more efficient to separate the resonance from the continuum 
states by increasing the attraction of the three-body force, and trace how the resonance
moves when the attraction of the three-body force is progressively released to the 
desired value. This is illustrated in Figs.~\ref{fig1}b and Figs.~\ref{fig1}c, where the red crosses 
indicate the position of the resonances when the strength of the three-body force is
put equal to $-12.1$ MeV in the $2^+$ case and 7 MeV in the $4^+$ case. 
The red arrows show how the resonances move when when decreasing the attraction.
Using this procedure the $2^+_3$ and $4^+_1$ resonances can be unambiguously identified. 

It is interesting to note that the states denoted as $4^+_1$ and $4^+_2$
in Fig.~\ref{fig1}c do actually cross when decreasing the repulsion of the three-body force
(red crosses in the figure), in such a way that eventually, for a sufficiently large
three-body attraction, the narrow resonance becomes 
the first $4^+$ state and the broad one becomes the second $4^+$ state. 
In other words, when decreasing the repulsion of the three-body force, the state denoted 
in Fig.~\ref{fig1}c as $4^+_2$ appears closer to the threshold than
the $4^+_1$ state.  It is important to keep this fact in mind 
in order to determine which of the two first $4^+$ states should be assigned to the first
$^{12}$C band, and which one to the second.

\subsection{$^{12}$C radii}

In order to get a feeling of the size of the states, we show in Table~\ref{tab3} the expectation value
$\langle \rho^2 \rangle^{1/2}$ for the same cases shown in Table~\ref{tab2}. This expectation value 
will be denoted as $\rho_{rms}$, and it 
is computed within the complex scaling frame, meaning that 
$\rho_{rms}^2=\langle \Psi |\rho^2 e^{i2\theta}|\Psi \rangle$, where $\Psi$ is the complex 
rotated wave function of the resonance and $\theta$ is the complex scaling angle used in the calculation 
\cite{alv07}. Therefore, the expectation value is not necessarily real even though the $\rho$-coordinate is.
With this value it is possible to obtain the root
mean square radius of each state ($r_{rms}$ in Table~\ref{tab3}), which is given by:
\begin{equation}
r_{rms}^2=\frac{1}{12}\langle \Psi | \rho^2 | \Psi \rangle e^{i2\theta}+R_\alpha^2,
\end{equation}
where $R_\alpha=1.47$ fm is the root mean square radius of the $\alpha$-particle. 
Again, due to the complex scaling calculation, the computed $r_{rms}$ of a resonance is in general 
a complex number.
As discussed in Ref.\cite{moi98}, the imaginary part of a computed observable (the energy is the most obvious
example) can be attempted interpreted as the uncertainty of the value given by the real part.
This quantity permits a fast comparison between the spatial extensions of the different states. 

As seen in the table, the results obtained with the Ali-Bodmer and Buck potentials are very
similar. It is remarkable that, on the one side,  the $0^+_1$ and $2^+_1$ states have similar sizes, both of
them in the vicinity of $\rho_{rms}\sim 6.8$ fm and $r_{rms}\sim 2.5$ fm, and, on the other side, the
same happens with the $0^+_2$ and $2^+_2$, resonances, with a value of 
$\rho_{rms}\sim 10$ fm and $r_{rms}\sim 3.4$ fm. This may suggest that these two sets of $0^+$ and $2^+$ 
states could correspond to two different rotational bands each of them with a reasonably 
well-``frozen'' structure. It is interesting to note that for the $4^+_1$ and $4^+_2$ states the 
computed values of 
$\rho_{rms}$ are precisely about 10 fm and 6 fm, respectively, which might indicate 
a crossing of the first and second $4^+$ states, in such a way that the $4^+_1$ state could belong 
to the second band and the $4^+_2$ state could belong to the first one. This is consistent with the
discussion in Fig.~\ref{fig1}c, where we showed that the $4^+_2$ state becomes actually the first $4^+$
and the $4^+_1$ state becomes the second $4^+$ when the repulsion in the three-body force is diminished.

Finally, a third band could be present containing the  $0^+_3$, $2^+_3$, and $4^+_3$ resonances. 
However, the values of $\rho_{rms}$ given in
Table~\ref{tab3} for these three states is not very stable (it ranges from 6.0 fm to 11.3 fm), and
furthermore, as seen in Table~\ref{tab2}, the experimental energy of the $0^+_3$ state is higher
than the one of the $2^+_3$ state, which makes it very unlikely that these two states belong to the same
rotational band.

\section{Electric quadrupole cross sections}

In this section we shall consider electric quadrupole cross sections
for transitions between the different $0^+$, $2^+$, and $4^+$ states
in $^{12}$C. These transitions involve bound states and resonances.
As already mentioned, in this work we do not treat the
resonances separately.  We just compute the wave functions for the discrete
continuum states obtained by imposing a box boundary condition. The
only role of the resonance energies obtained in the previous section
is to provide information about where to put the energy windows when
computing the cross sections.

In general, the cross section, as a function of the incident energy,
is given by Eq.(\ref{totcs}), where the summation over $j$ is
restricted to the states within the chosen final energy window.  For
clean isolated resonance peaks it is easy to choose a meaningful, and
rather well defined, window around the peak.  For more complicated
energy dependent cross sections this could easily be more ambiguous.
However, this would only reflect the more complicated physical nature
resulting from overlapping and perhaps interfering resonances.
Variation of choices of windows and subsequent analyses are then
related to reduction of contributions to properties of individual
resonances.  Thus the ambiguity in the choice of the window still remains but
now containing information about the underlying physics

We shall here define the windows as $E_R\pm\Gamma_R$, where $E_R$ and
$\Gamma_R$ are the energy and the width of the resonance in the
initial or final state (we shall use the computed energies and widths
given in Table~\ref{tab2}). Obviously, when the final state is a bound
state the corresponding width is zero, and the summation over $j$ in
Eq.(\ref{totcs}) disappears and the wave function in the right part of
the matrix element refers to the bound-state wave function.

The continuum states have been discretized by imposing a box boundary
condition to the radial solutions $f_n(\rho)$ in the set of
Eqs.(\ref{eq9}). In particular, a box size of 200 fm has been used,
which amounts to an average energy separation of about 0.03 MeV
between the lower-lying continuum states.  This means that for an
energy window with a width in the scale of MeV we will have a
significant amount of discrete continuum states within the final
energy window. However, this procedure can not be applied to
transitions into the Hoyle state. This resonance is extremely narrow
(about $8$~eV), which implies that only a huge box could provide a
significant number of discrete continuum states within a final energy
window of only a few eV wide. 

Discretization of the continuum by use of such a huge box of perhaps
$30000$~fm compared to the 200~fm used in this work is clearly
meaningless within the present numerical approach.  The reasons are
the huge number of necessary partial waves, the increasing basis size
for each of the corresponding components, and the coupling of all the
potentials due to the long-range Coulomb force.  Therefore the
transitions into the Hoyle state will be treated as ordinary
transitions into a bound state.  On the other hand this is a very
accurate approximation for such a small width.

\subsection{$2^+\rightarrow 0^+$ transitions}

\begin{figure}
\epsfig{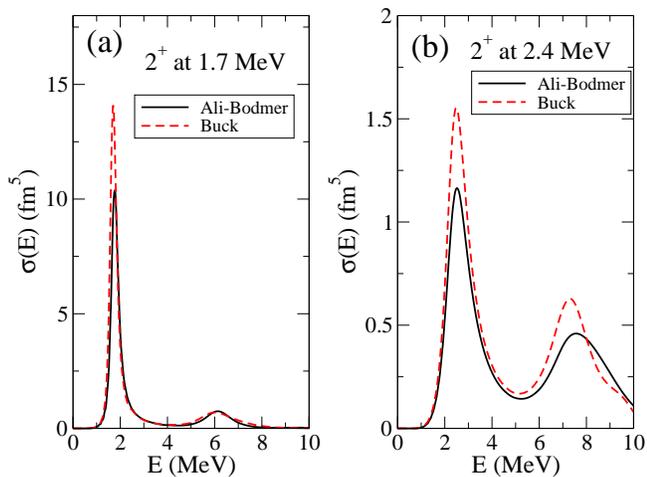}
\caption{(Color online)
Cross section (Eq.(\ref{totcs})) for the $2^+ \rightarrow 0^+_1$ transition in the $\alpha+\alpha+\alpha
\rightarrow \mbox{$^{12}$C}+\gamma$ reaction as a function of the three-body energy. 
The results with the Ali-Bodmer potential (solid curves) and the Buck potential (dashed curves) are shown. 
The (a) and (b) panels correspond to the calculations placing the $2^+$ resonance at 1.7 MeV and 
2.4 MeV, respectively.  }
\label{fig2}
\end{figure}

In Fig.\ref{fig2} we show the cross section for the
$\alpha+\alpha+\alpha \rightarrow \mbox{$^{12}$C}+\gamma$ reaction for
a transition between the continuum $2^+$ states and the ground state
in $^{12}$C.  The three-body force for the $0^+$ state has been chosen
to reproduce the experimental separation energy of $^{12}$C into three
alphas (see Table~\ref{tab2}). The figure shows the results obtained
with the Ali-Bodmer potential (solid curves) and the Buck potential
(dashed curves).  Two different (structureless) three-body potentials
have been used to place the $2^+$ resonance at different energies.  

In panel (a) the $2^+$ energy is chosen to reproduce the measured
binding energy of the bound $2_1^+$ state.  This implies that the
first resonance, $2_2^+$, is found at about 1.7 MeV, which gives rise
to the pronounced peak in the cross section around this energy.  As we
can see, the calculation using the Buck potential produces a taller
peak than with the Ali-Bodmer potential. This is related to
the fact that the computed $2^+_2$ state is clearly narrower when the
Buck potential is used.
By fitting these peaks with the
Breit-Wigner shape in Eq.(\ref{bw2}) we extract the value of
$\Gamma_\gamma$ for the $2^+_2 \rightarrow 0^+_1$ reaction, which 
for the Ali-Bodmer and Buck potentials is found to be  
205~meV and 115~meV, respectively.  We emphasize that, as seen from Eq.(\ref{bw2}), 
in the center of the Breit-Wigner shape ($E=E_R$)
the value of the cross section is proportional to
$\Gamma_\gamma/\Gamma_R$ (assuming that $\Gamma_\gamma \ll \Gamma_R$),
and therefore different values of the cross section in the peak does
not necessarily imply different values of the $\Gamma_\gamma$-width,
or, in other words,
only for similar values of $\Gamma_R$ the difference in the peaks of
the cross section is directly translated into a difference in the 
$\Gamma_\gamma$-values.
The computed $\Gamma_\gamma$-widths are collected in Table~\ref{tab4}.
Note that the presence of the $2^+_3$ resonance at about 4 MeV with 
the Ali-Bodmer potential or 5 MeV with the Buck potential has no
visible effect on the cross section. In fact, the small peak
observed at around 6 MeV is basically produced by the broad $2^+_4$ 
resonance, which with both potentials appears close to 6 MeV, as shown
for the Buck potential by the red crosses in Fig.~\ref{fig1}b.

\begin{table}
\begin{center}
\caption{$\Gamma_\gamma$-widths, in meV, for the resonance-to-resonance transitions obtained after fitting 
the peaks in the cross sections for the different reactions with the Breit-Wigner shape given in 
Eq.(\ref{bw2}). For the reactions where the $2^+_2$ resonance enters, the two resonance energies, 1.7 MeV 
and 2.4 MeV, have been considered.}
\label{tab4}
\begin{ruledtabular}
\begin{tabular}{|c|cc|cc|}
               & \multicolumn{2}{c|}{Ali-Bodmer} & \multicolumn{2}{c|}{Buck} \\ 
      $E_{2^+}=$         &   1.7 MeV  &  2.4 MeV  &  1.7 MeV  &  2.4 MeV  \\ \hline
$2^+_2 \rightarrow 0^+_1$ & 205   & 160 &  115  & 175  \\
$2^+_2 \rightarrow 0^+_2$ & 1.0   & 5.9 &  0.4  & 4.5  \\
$2^+_3 \rightarrow 0^+_1$ & ---   & 1950&  ---  & 4300  \\
$2^+_3 \rightarrow 0^+_2$ & 0.7   &  80 &  6.5  & 190  \\
$2^+_2 \rightarrow 2^+_1$ & 6.4   &  22 &  4.0  & 22    \\
$2^+_3 \rightarrow 2^+_1$ &  20   & 320 &  88   & 950    \\
$4^+_1 \rightarrow 2^+_2$ &  130  & 44 &  118   & 57  \\
$4^+_2 \rightarrow 2^+_2$ &  36   & 25 &   26   & 18   \\
$4^+_1 \rightarrow 2^+_1$ &    \multicolumn{2}{c|}{50}  &   \multicolumn{2}{c|}{50}  \\      
$4^+_2 \rightarrow 2^+_1$ &    \multicolumn{2}{c|}{635} &   \multicolumn{2}{c|}{860} \\      
$4^+_3 \rightarrow 2^+_1$ &    \multicolumn{2}{c|}{3100} &   \multicolumn{2}{c|}{3500} \\      
\end{tabular}
\end{ruledtabular}
\end{center}
\end{table}

In the second calculation (Fig.\ref{fig2}b) the three-body force for the
$2^+$ states has been chosen such that the lowest $2^+$ resonance is placed at
about 2.4 MeV, in closer agreement to the experimental value given in
Ref.\cite{zim13}. The peaks in the cross sections are therefore
shifted towards higher energies. As we
can see, the new peaks are now clearly broader and almost a 
factor of 10 lower than before. Again, the calculation with the Buck potential 
gives rise to a taller peak than when the Ali-Bodmer potential is used. The
$\Gamma_\gamma$-widths obtained with the two
potentials are in this case very similar, 160~meV and 175~meV,
respectively. In this case the second peak in the cross section
located around 7 MeV is produced by the $2^+_3$ resonance. 
From this peak we can also extract the $\Gamma_\gamma$ widths for the
$2^+_3 \rightarrow 0^+_1$ reaction, which for completeness are also given in 
Table~\ref{tab4}. 

\begin{figure}
\epsfig{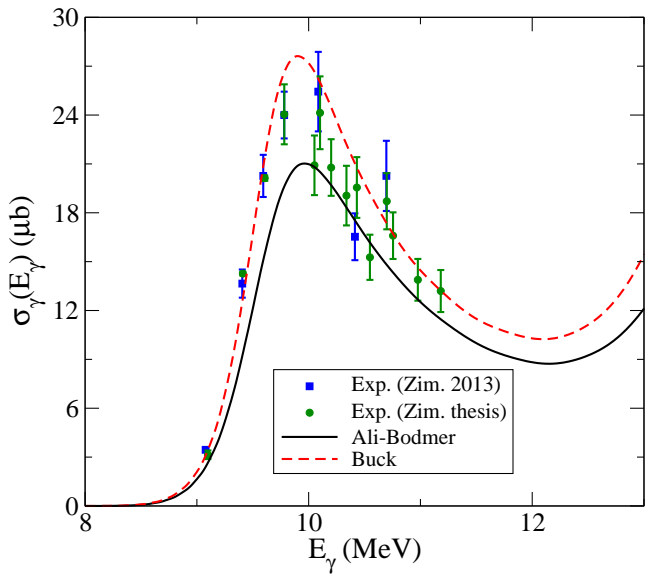}
\caption{(Color online)
Photo-dissociation cross section for the $\mbox{$^{12}$C}+\gamma \rightarrow \alpha+\alpha+\alpha$ reaction
between the $0^+$ ground state and the continuum $2^+$ states as a function of the photon energy. 
The experimental curve has  been multiplied by a factor of 3. The experimental data are from 
Refs.\cite{zim13} 
(squares) and \cite{zim13b} (circles). Only the results with the $2^+$ resonance at 2.4 MeV
are shown for both, the Ali-Bodmer (solid) and Buck (dashed) potentials. }
\label{fig3}
\end{figure}

In Ref.\cite{zim13} the $\Gamma_\gamma$-width of the $2^+_2$ resonance
in $^{12}$C has been extracted from the photo-dissociation cross
section of the $0^+_1$ ground state, and it has been found to be
$60\pm10$~meV, which is roughly a factor of 3 smaller than the values
obtained in this work when the 2$^+$ resonance is placed at 2.4
MeV (which is the energy that agrees better with the one given
in the same reference). In Ref.\cite{zim13b}, where the same experiment is analyzed, the
$\Gamma_\gamma$-width of the $2^+_2$ state is quoted to be $135\pm 14$
meV, which agrees better with the results of our computation. However,
as mentioned above, the relevant quantity when fitting the data with a
Breit-Wigner shape is $\Gamma_\gamma/\Gamma_R$, which in both,
Ref.\cite{zim13} and Ref.\cite{zim13b}, is of about $6.5\cdot 10^{-8}$. This
value is roughly a factor of 3 smaller than the value we have obtained
with the $2^+_2$ resonance at about 2.4 MeV ($\Gamma_\gamma/\Gamma_R \approx 17\cdot 10^{-8}$). 
Therefore the photo-dissociation cross section for the $\mbox{$^{12}$C}+\gamma
\rightarrow \alpha+\alpha+\alpha$ shown in Refs.\cite{zim13,zim13b}
should also be roughly a factor of 3 smaller than the one obtained in
this work.  

The photo-dissociation cross section can be extracted from the cross
sections shown in Fig.\ref{fig2} simply by using the relation in
Eq.(\ref{eq3}). The result is shown in Fig.\ref{fig3}, where the
experimental data (multiplied by a factor of 3) are the ones given in 
Refs.\cite{zim13} (squares)
and \cite{zim13b} (circles) and the curves are our calculations with
the Buck and Ali-Bodmer potentials (and the energy of the $2^+_2$
resonance at about 2.4 MeV). As we can see, the shape of the experimental cross section is
well reproduced. The disagreement on the absolute value of the cross
section may be related to the fact that only a fraction of the $0^+$
ground state in $^{12}$C corresponds to the three-alpha structure used
in this work.  Another possibility is that improvements of the
experimental analyses with the available conflicting results
\cite{zim13,zim13b} also would change the normalization.

\begin{figure}
\epsfig{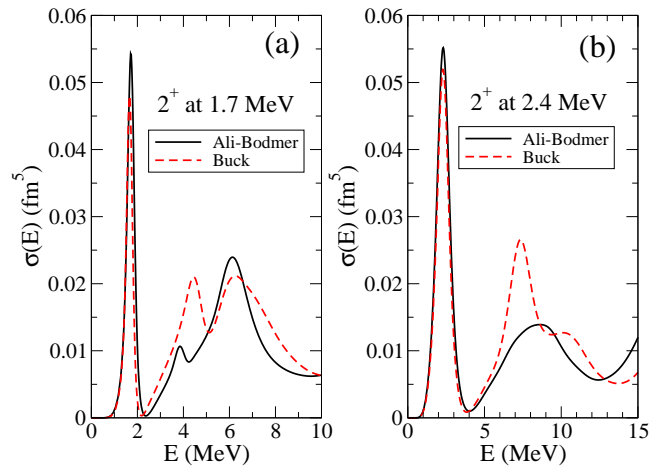}
\caption{(Color online)
The same as Fig.\ref{fig2} for 
the $2^+ \rightarrow 0^+_2$ transition in the $\alpha+\alpha+\alpha
\rightarrow \mbox{$^{12}$C}+\gamma$ reaction. }
\label{fig4}
\end{figure}

In Fig.\ref{fig4} we show the same as in Fig.\ref{fig2} for
transitions from the $2^+$ continuum states in $^{12}$C into the Hoyle
state (treated as a bound state). Again, the (a) and (b) panels
show the result when the $2^+_2$ resonance is at 1.7 MeV and 2.4 MeV,
respectively. The computed cross sections are clearly smaller than for
transitions into the $0^+$ ground state.  Together with the main peak
produced by the $2^+_2$ resonance, several other peaks produced by higher
$2^+$ states are clearly seen. In particular, in Fig.\ref{fig4}a a
small peak can be seen at an energy of around 4 MeV for the Ali-Bodmer
potential and 5 MeV for the Buck potential. These two peaks are
produced by the $2^+_3$ resonance given in Table~\ref{tab2}. An
additional peak produced by the $2^+_4$ resonance at about 6 MeV is also 
seen in the cross section. The computed $\Gamma_\gamma$-widths obtained 
after fitting the lower peak in Fig.\ref{fig4}a with the Breit-Wigner shape 
in Eq.(\ref{bw2}) are 1.0 meV and 0.4 meV with the Ali-Bodmer and Buck 
potentials, respectively. 

When the $2^+_2$ resonance is at 2.4 MeV (Fig.\ref{fig4}b)
the $2^+_3$ state becomes very broad and moves up in energy,
giving rise to the
corresponding bumps in the cross section observed in Fig.\ref{fig4}b. In the case of 
the Buck potential a bump produced by the $2^+_4$ state at about 11 MeV 
(Fig.\ref{fig1}b) is also seen. The computed $\Gamma_\gamma$ widths for the
transition from the $2^+_2$ state to the Hoyle state are in this case
5.9 meV and 4.5 meV with the Ali-Bodmer and the Buck potentials,
respectively. For completeness, we also give in Table~\ref{tab4} the computed 
values of $\Gamma_\gamma$ for the $2^+_3 \rightarrow 0^+_2$ transition.

\subsection{$4^+\rightarrow 2^+$ transitions}

\begin{figure}
\epsfig{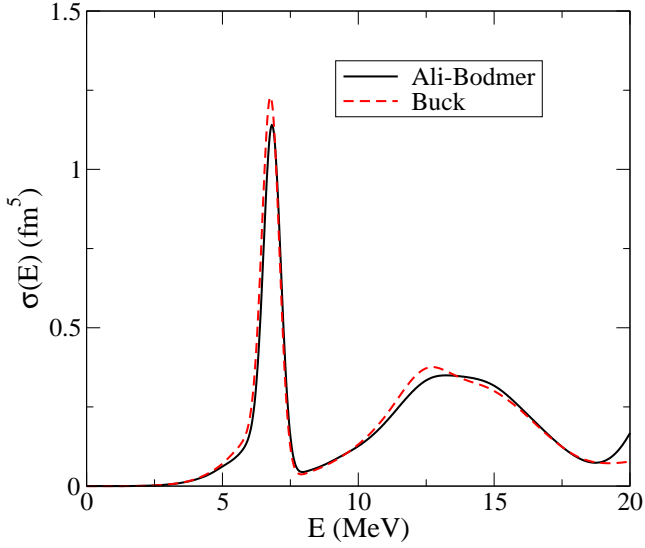}
\caption{(Color online)
The same as Fig.\ref{fig2} for 
the $4^+ \rightarrow 2^+_1$ transition in the $\alpha+\alpha+\alpha
\rightarrow \mbox{$^{12}$C}+\gamma$ reaction. }
\label{fig5}
\end{figure}

\begin{figure}
\epsfig{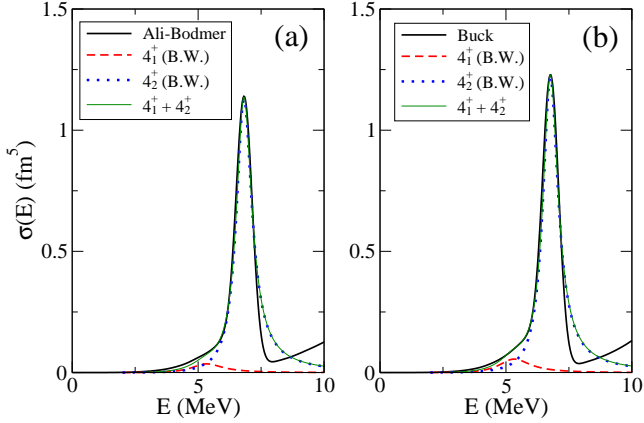}
\caption{(Color online)
Fit of the sharp peak in the cross section for the $4^+ \rightarrow 2^+_1$ transition 
by means of two Breit-Wigner functions with the parameters corresponding to
the $4^+_1$ and $4^+_2$ resonances given in Table~\ref{tab2}. Panels (a) and (b) 
show the fit when the Ali-Bodmer and Buck potentials, respectively, are used. The thick-solid curve
curve is the computed cross section as shown in Fig.\ref{fig5}. The dashed and dotted
curves are the Breit-Wigner functions corresponding to the $4^+_1$ and $4^+_2$ resonances,
respectively, whose sum is shown by the thin-solid curve.}
\label{fig6}
\end{figure}

In Fig.\ref{fig5} we show the cross section for the $4^+\rightarrow 2^+_1$ transition. The three-body
force in the $2^+$ states is such that the binding energy of the $2^+_1$ bound state matches the
experimental value. The two potentials, Ali-Bodmer and Buck, give very similar results with a very
sharp peak at 6.8 MeV, which is actually produced by the $4^+_2$ resonance. The effect of the 
broad 4$^+_1$ state can be seen in the little shoulder that appears in the cross section around 5 MeV.
This is more clearly seen in Fig.\ref{fig6}, where we fit for the Ali-Bodmer (panel (a)) and
Buck (panel (b)) potentials the cross section peak as a linear combination of two Breit-Wigner 
functions with the parameters for the $4^+_1$ (dashed curves) and $4^+_2$ (dotted curves) resonances 
given in Table~\ref{tab2}. The sum of the two Breit-Wigner curves is shown by the thin-solid curves, 
which match pretty well the peak of the computed cross sections (thick-solid curves). 
The coefficients in this linear combination give directly the
$\Gamma_\gamma$ widths for the $4^+_1\rightarrow 2^+_1$ and $4^+_2\rightarrow 2^+_1$ transitions,
which are, with the Ali-Bodmer and Buck potentials, 50 meV and 80 meV for the first transition, and
635 meV and 855 meV for the second transition, respectively. The broad peak observed at about 13 MeV 
is mainly produced by the 4$^+_3$ resonance, which has very similar properties with the two 
potentials. The $\Gamma_\gamma$ values for the $4^+_3 \rightarrow 2^+_1$ transitions are also 
given in Table~\ref{tab4}.

\begin{figure}
\epsfig{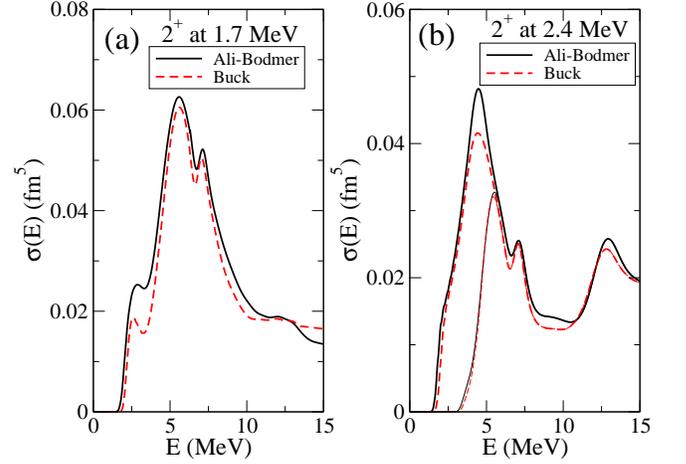}
\caption{(Color online)
The same as Fig.\ref{fig2} for 
the $4^+ \rightarrow 2^+_2$ transition in the $\alpha+\alpha+\alpha
\rightarrow \mbox{$^{12}$C}+\gamma$ reaction. In panel (b) the thin-solid
and thin-dashed curves are the cross sections for the Ali-Bodmer and Buck 
potentials, respectively, using a cutoff for the photon energy of 2 MeV.}
\label{fig7}
\end{figure}

\begin{figure}
\epsfig{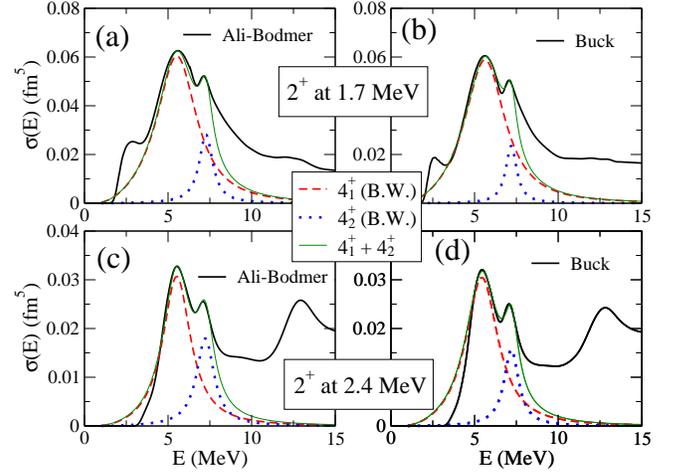}
\caption{(Color online)
Fit of the cross section for the $4^+ \rightarrow 2^+_2$ transition
by means of two Breit-Wigner functions with the parameters corresponding to
the $4^+_1$ and $4^+_2$ resonances given in Table~\ref{tab2}. Panels (a) and (b)
show the fit when the Ali-Bodmer and Buck potentials, respectively, are used and the 
$2^+$ resonance is found at 1.7 MeV. Panels (c) and (d) show the same as (a) and (b) 
when the $2^+$ resonance energy is 2.4 MeV. The thick-solid curve
is the computed cross section as shown in Fig.\ref{fig7} (with the photon energy
cutoff in panels (c) and (d)). The dashed and dotted
curves are the Breit-Wigner functions corresponding to the $4^+_1$ and $4^+_2$ resonances,
respectively, whose sum is shown by the thin-solid curve.  }
\label{fig8}
\end{figure}

In Fig.\ref{fig7} we show the cross sections for the $4^+\rightarrow
2^+_2$ transitions. Again, two possible energies, 1.7 MeV and 2.4 MeV,
have been considered for the $2^+_2$ resonance (panels (a) and (b),
respectively). This is a transition into a resonant state, which implies
that the summation over $j$ in Eq.(\ref{totcs}) is made over the 
discrete continuum final states within the energy window $E_R\pm\Gamma_R$, 
where $E_R$ and $\Gamma_R$ are the energy and width of the final resonance. 
Therefore, the value of the cross section will be fully determined by the arbitrary
choice for the width of the final energy window. 

The calculation of the continuum-to-continuum cross section
presents also the numerical complication of the so-called infrared catastrophe 
(see Ref.\cite{gar12}), which appears in the region of small $E_\gamma$ values
($E$ close to $E^\prime$) 
due to the $1/E_\gamma$ dependence of the bremsstrahlung cross section 
at small photon energies. Therefore, the cross sections shown in Fig.\ref{fig7}
by the thick-solid (Ali-Bodmer) and thick-dashed (Buck) curves
will be contaminated by this effect, also called soft-photon contribution, for initial 
energies ($E$) in the vicinity
of the final energy window (centered at $\sim 1.75$ MeV in panel (a) and
at $\sim 2.4$ MeV in panel (b)). A fully relativistic treatment of the bremsstrahlung 
cross section would correct this anomaly \cite{gre01}. 

In Fig.\ref{fig7}a the effect
of the soft-photon contribution is seen in the little shoulder observed at about 2 MeV.
This peak is therefore unphysical and it does not correspond to any resonance 
in the initial state. The other two peaks, at about 5.5 MeV and 7.2 MeV, correspond
to the $4^+_1$ and $4^+_2$ states, although the large width of the $4^+_1$ resonance gives 
rise to a large interference between the two lowest $4^+$ states.

In the case shown in Fig.\ref{fig7}b ($2^+_2$ resonance at $\sim 2.4$ MeV) the final energy 
window reaches a value of around 3.5 MeV. For this reason the soft-photon peak and the
one corresponding to the broad $4^+_1$ resonance can not be distinguished in this case. 
Usually the soft-photon contribution is removed by introducing a cutoff in the 
photon energy. In Fig.\ref{fig7}b the thin-solid and thin-dashed curves are the cross sections 
with the Ali-Bodmer and Buck potentials, respectively, when a cutoff of 2 MeV is used for 
$E_\gamma$. This cutoff has been chosen in order to place the $4^+_1$ peak at an energy of 
$\sim 5.5$ MeV, similar to the one observed in panel (a), that we know from the value of the 
$4^+_1$ energy should be the correct position. Although the choice of the cutoff value is 
to some extent arbitrary, it should not be much bigger than 2 MeV, since the energy separation between the
4$^+_1$ and the $2^+_2$ resonances in Fig.\ref{fig7}b is of about 3 MeV. For this reason, 
the transition strengths obtained from the lowest peak in Fig.\ref{fig7}b (thin curves) for the 
$4^+_1 \rightarrow 2^+_2$ process should be taken as a minimum value. A choice of the
cutoff energy smaller than 2 MeV would increase the strength. The second peak at $\sim 7.1$ MeV 
is produced by the $4^+_2$ state. The position and height of this peak is
not affected by the cutoff in the photon energy, but, again, it contains an important interference 
from the $4^+_1$ resonance.

In order to extract the $\Gamma_\gamma$-values for the $4^+_1 \rightarrow 2^+_2$
and $4^+_2 \rightarrow 2^+_2$ transitions, and due to the large interference
between the two lowest $4^+$ resonances, it is convenient to proceed as
discussed in Fig.\ref{fig6}, and make a simultaneous fit of the $4^+_1$
and $4^+_2$ peaks in the cross sections by means of Eq.(\ref{bw2}). These fits
are shown in Fig.\ref{fig8}. The panels (a) and (b) 
are the calculations with the Ali-Bodmer and Buck potentials for the $2^+$ state 
at 1.7 MeV, and panels (c) and (d) are the same calculations with the $2^+$ state 
at 2.4 MeV. In these two last panels the fit has been made to the cross sections 
obtained after removing the soft-photon contribution by means of the cutoff in 
$E_\gamma$ (thin curves in Fig.\ref{fig7}b). The Breit-Wigner curves 
corresponding to the $4^+_1$ and $4^+_2$ states are shown by the dashed
and dotted curves, respectively, and they have been obtained using 
$\Gamma_\gamma$-values equal to 130 meV and 36 meV, respectively, in panel (a),
118 meV and 26 meV in panel (b), 44 meV and 25 meV in panel (c), and 57 meV 
and 18 meV in panel (d). The sum of the two Breit-Wigner fits is given by the
thin-solid curves, which clearly disagree with the computed total cross section
(thick-solid curves) for large energies. This is due the fact neither the contribution 
from the wide $4^+_3$ state given in Table~\ref{tab2} nor the contribution
form the continuum background have been included in fit.  

\begin{figure}
\epsfig{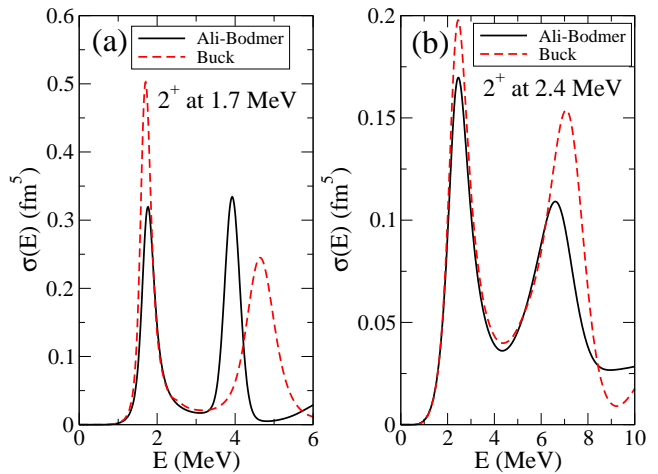}
\caption{(Color online)
The same as Fig.\ref{fig2} for 
the $2^+ \rightarrow 2^+_1$ transition in the $\alpha+\alpha+\alpha
\rightarrow \mbox{$^{12}$C}+\gamma$ reaction. }
\label{fig9}
\end{figure}

Finally, we show in Fig.\ref{fig9} the corresponding  cross sections for the $2^+\rightarrow 2^+_1$ 
transitions. The peaks produced by the $2^+_2$ and $2^+_3$ resonances are clearly seen in 
all the cases. The $\Gamma_\gamma$-widths for the $2^+_2\rightarrow 2^+_1$ and
$2^+_3\rightarrow 2^+_1$ transitions
with the two potentials and the two different energies of the $2^+_2$ resonance considered in 
this work are again given in Table~\ref{tab4}.

\section{${\cal B}^{(E2)}$ transition strengths}

\begin{table*}
\begin{center}
\caption{${\cal B}(E2)$ transition strengths in $e^2$fm$^4$ for different transitions between $^{12}$C
states. For transitions from continuum states, ${\cal B}_\gamma$ and ${\cal B}_\sigma$ denote the 
strengths obtained from Eq.(\ref{gamma}) and from the total (integrated) cross section, respectively.
For the reactions involving the $2^+_2$ resonance the two cases, with the resonance energy
at 1.7 MeV and 2.4 MeV, have been considered. 
The second, third, and fourth columns give the known experimental values, and the results obtained
with the microscopic cluster model calculation described in Ref.\cite{che07}
(labeled as $\alpha$-cluster) and with the Antisymmetrized Molecular Dynamics (AMD) method. 
The corresponding references are given in the table.}
\label{tab5}
\begin{ruledtabular}
\begin{tabular}{|c|ccc|cccc|cccc|}
Transition  &  Exp. & $\alpha$-cluster &  AMD &\multicolumn{4}{c|}{Ali-Bodmer}  & \multicolumn{4}{c|}{Buck} \\ \hline
$2^+_1 \rightarrow 0^+_1$ &  $7.6\pm0.4^{(a)}$& 9.16$^{(a)}$ & 8.4$^{(d)}$  &\multicolumn{4}{c|}{10.2} & \multicolumn{4}{c|}{9.9}  \\
$4^+_2 \rightarrow 2^+_1$ &           &      & 15.8$^{(d)}$  &\multicolumn{2}{c}{${\cal B}_\gamma=9.0$}  & \multicolumn{2}{c|}{${\cal B}_\sigma=8.2$} & \multicolumn{2}{c}{${\cal B}_\gamma=12.5$}  &  \multicolumn{2}{c|}{${\cal B}_\sigma=9.9$} \\ \hline
   &   &    &    & \multicolumn{2}{c}{$E_{2^+}$=1.7 MeV} & \multicolumn{2}{c|}{$E_{2^+}$=2.4 MeV} &
             \multicolumn{2}{c}{$E_{2^+}$=1.7 MeV} & \multicolumn{2}{c|}{$E_{2^+}$=2.4 MeV}   \\
   &   &    &    &  ${\cal B}_\gamma$ &   ${\cal B}_\sigma$ &  ${\cal B}_\gamma$ &   ${\cal B}_\sigma$ &  ${\cal B}_\gamma$ &  ${\cal B}_\sigma$ &   ${\cal B}_\gamma$ &   ${\cal B}_\sigma$  \\
$2^+_2 \rightarrow 0^+_2$ &           &      & 102$^{(d)}$ &  230 & 175  & 170 & 140 & 120  & 135 & 158 & 136  \\
$4^+_1 \rightarrow 2^+_2$ &           &      & 595$^{(d)}$ &  170  & 135  & 153 & 130 &  150  & 130 & 220 & 145 \\ \hline
$2^+_1 \rightarrow 0^+_2$ &  $2.6\pm0.4^{(a)}$& 0.84$^{(a)}$ & 5.1$^{(d)}$  &\multicolumn{4}{c|}{0.91} & \multicolumn{4}{c|}{0.96} \\
$4^+_1 \rightarrow 2^+_1$ &           &      &               &\multicolumn{2}{c}{${\cal B}_\gamma=1.2$}  & \multicolumn{2}{c|}{${\cal B}_\sigma= 0.6$} & \multicolumn{2}{c}{${\cal B}_\gamma=1.2$}  &  \multicolumn{2}{c|}{${\cal B}_\sigma=1.0$}  \\ 
   &   &    &    & \multicolumn{2}{c}{$E_{2^+}$=1.7 MeV} & \multicolumn{2}{c|}{$E_{2^+}$=2.4 MeV} &
             \multicolumn{2}{c}{$E_{2^+}$=1.7 MeV} & \multicolumn{2}{c|}{$E_{2^+}$=2.4 MeV}   \\
   &   &    &    &  ${\cal B}_\gamma$ &   ${\cal B}_\sigma$ &  ${\cal B}_\gamma$ &   ${\cal B}_\sigma$ &  ${\cal B}_\gamma$ &  ${\cal B}_\sigma$ &   ${\cal B}_\gamma$ &   ${\cal B}_\sigma$   \\
$2^+_2 \rightarrow 0^+_1$ &$0.73\pm 0.13^{(b)}$& 1.99$^{(a)}$ & 0.4$^{(d)}$ & 4.1 & 3.1 & 2.0 & 1.9 & 2.4 & 2.8 & 2.3 & 2.2  \\
                          &$1.57\pm 0.13^{(c)}$&      &     &      &      &      &      &       &      &      &       \\
$2^+_2 \rightarrow 2^+_1$ &           &      &            &  3.6 &  2.9 &  5.5 &  5.2 & 2.5  & 2.9 & 5.5 & 5.4    \\
$4^+_2 \rightarrow 2^+_2$ &           &      & 7.5$^{(d)}$ & 8.5 &  4.0 &  9.8 &  5.0 & 6.5  & 3.5 & 7.5 & 4.7 \\
\end{tabular}
$(a)$ Ref.\cite{che07}; $(b)$ Ref.\cite{zim13}; $(c)$ Ref.\cite{zim13b}; $(d)$ Ref.\cite{cuo13}
\end{ruledtabular}
\end{center}
\end{table*}

The photo-emission processes from a bound state into other bound
states or into the continuum is described unambiguously by the
transition strength. This is not true any more for continuum to
continuum transitions even when rather well-defined resonance peaks
are present in the corresponding cross sections.  These transition
strengths are decisive indicators for structure similarities between
excited states, where prominent examples are collective rotational or
vibrational states built on the same intrinsic configuration.  In our
present case of three alpha-particles such collective rotations have
been suggested many times. This necessarily involves continuum
structures with all the related ambiguities.  Still, we want to know
if collective rotations is a reasonable description of some of the
states in the $^{12}$C-spectrum.  We therefore first discuss the
detailed numerical results from our three-body model, and in the
following subsection we relate to the simplest possible rotational
model.

\subsection{Numerical results}

In Sect.~\ref{trans} we described the two methods used in this work to
extract the transition strength for reactions involving continuum
states. In the first method we integrate the cross section (Eq.(\ref{eq6}))
over two energy windows chosen around the initial and final
resonance energies. In this work we have taken the windows
as $E_R \pm \Gamma_R$ where $\Gamma_R$ is the width of the resonance.
This procedure is equivalent to integration of the cross sections computed in 
the previous section over the initial energy $E$ under the peaks
corresponding to the initial resonant state. The
total cross section obtained in this way is divided by the constants
multiplying the transition strength (see Eqs.(\ref{eq4}) and
(\ref{eq6})).  These transition strengths will be denoted by ${\cal
  B}_\sigma$. The application of this method requires well-separated
peaks in the cross sections for each of the resonances, in such a way
that the integration under a given peak contains very little contamination
from the interference with other states. However, this is not always
the case in our calculations. In particular, the $4^+_1$ and $4^+_2$ states
are very close to each other, giving rise to a large interference
between them, as seen in Figs.\ref{fig6} and \ref{fig8}. For this
reason, for transitions having the $4^+_1$ or $4^+_2$ resonances as initial
state, the integration of the cross section over the initial energy window 
will be made considering not the computed cross section (thick solid curves
in Figs.\ref{fig6} and \ref{fig8}), but the Breit-Wigner curves corresponding
to each of the resonances (dashed curves for the $4^+_1$ state and dotted 
curve for the $4^+_2$ state). In the second method we make use of Eq.(\ref{gamma}), 
where $\Gamma_\gamma$ has been extracted after fitting the computed cross section with the
Breit-Wigner shape in Eq.(\ref{bw2}).  The values of $\Gamma_ \gamma$
have been given in Table~\ref{tab4}.  The transitions strengths
computed in this way will be denoted by ${\cal B}_\gamma$.  

The first two lines in Table~\ref{tab5} show the computed strengths
for the $2^+_1 \rightarrow 0^+_1$ and $4^+_2 \rightarrow 2^+_1$
transitions, which are tempting to associate with transitions between
states belonging to the first rotational band in $^{12}$C. In here 
we have taken into account that, as suggested by the computed values
of $\rho_{rms}$ (see Table~\ref{tab3}) and the
behavior of the $4^+_1$ and $4^+_2$ resonances when modifying the three-body
force (Fig.\ref{fig1}), the $4^+_2$ state is expected to belong to
the first band and the $4^+_1$ state to the second.  The first
transition in the table involves only bound states, and the results obtained with
the Ali-Bodmer and Buck potentials are similar, 10.2 $e^2$fm$^4$ and
9.9 $e^2$fm$^4$ which are in good agreement with the result in
Ref.\cite{che07} from a microscopic $\alpha$-cluster model calculation. These
values are a bit bigger than the one obtained in Ref.\cite{cuo13}
where the microscopic Antisymmetrized Molecular Dynamics (AMD) method
was used.  The results are also bigger than the experimental value
\cite{che07} which most temptingly can be attributed to a better match
between the $3\alpha$ wave functions than nature prescribes.

For the $4^+_2 \rightarrow 2^+_1$ transition the computed 
${\cal B}_\gamma$-strengths agree slightly better with the AMD calculation than
the ${\cal B}_\sigma$-values, which are a bit smaller. In comparison
to other model calculations it is important to know the conceptual
difference between our continuum calculation and these bound-state
treatments.  When a resonance is located in the continuum with a
substantial width it is very likely that the transition strength is
reduced due to a smaller overlap of wave functions than by assuming
one bound-state like resonance wave function.  It is also important to
keep in mind that the choice of the energy windows when computing
${\cal B}_\sigma$ is arbitrary, and a small increase in such windows
will give rise to values of ${\cal B}_\sigma$  closer to ${\cal B}_\gamma$.

The next two lines in the table give the $E2$-transition strengths
between the states in the second sequence of energies,
i.e., the $2^+_2 \rightarrow 0^+_2$ and $4^+_1 \rightarrow 2^+_2$
transitions.  In order to see the dependence on the energy of the
$2^+_2$ resonance, we show, as in Table~\ref{tab4}, the results
corresponding to a $2^+_2$ energy of 1.7 MeV and 2.4 MeV. 

For the $2^+_2 \rightarrow 0^+_2$ transition, when the $2^+_2$ resonance
is at 1.7 MeV, the results with the Buck potential are clearly smaller than 
the values obtained with the Ali-Bodmer potential. This is particularly true
for ${\cal B}_\gamma$, where the difference is of almost a factor of 2.
This is due to the fact that the width of the $2^+_2$ resonance (at 1.7
MeV) with the Buck potential is around half the width of the one with
the Ali-Bodmer potential (see Table~\ref{tab2}). As seen in
Fig.\ref{fig4}a, the height of the cross section peak corresponding to
this resonance is similar for both potentials.  This implies that
$\Gamma_\gamma/\Gamma_R$ is also similar in both cases. Therefore, a
$\Gamma_R$-value about a factor of 2 smaller gives rise to a
$\Gamma_\gamma$-value also about a factor of 2 smaller, and
consequently, as seen in Eq.(\ref{bw2}), a transition strength also
about a factor of 2 smaller.
For a $2^+_2$ energy of 2.4 MeV, which is in better agreement with the
experimental value given in \cite{zim13,zim13b}, the computed
strengths with the Ali-Bodmer and Buck potentials are now closer to each
other (the difference between the corresponding $\Gamma_R$ values is
now small). Also, the ${\cal B}_\gamma$ and ${\cal B}_\sigma$ values
are reasonably consistent with each other, and they are clearly bigger than
the AMD result ($102$ $e^2\mbox{fm}^4$) given in Ref.\cite{cuo13}. In
any case this difference is relatively unimportant compared to the
sensitivity of the computed transition strengths on the methods
used. 

For the $4^+_1 \rightarrow 2^+_2$ transition the results obtained
with the two potentials are consistent with each other, even if
${\cal B}_\gamma$ changes from 153 $e^2$fm$^4$ to 220 $e^2$fm$^4$
when the 2$^+_2$ state is at 2.4 MeV. One has to take into account that, 
together with the inherent
uncertainties in the determination of ${\cal B}_\gamma$ and ${\cal B}_\sigma$,
for reactions involving the
4$^+_1$ and $4^+_2$ states we have to deal with the additional
uncertainty arising from the interference between the two resonances
and the soft-photon contribution. For this reason, the agreement
between the ${\cal B}_\gamma$ and ${\cal B}_\sigma$ values seen
in Table~\ref{tab5} for the  $4^+_1 \rightarrow 2^+_2$ reaction
can be considered quite acceptable. In any case, all the results
given in the Table for this reaction are significantly smaller than the 
result given in Ref.\cite{cuo13}.

In the lower part of Table~\ref{tab5} we show the computed transition strengths between states
belonging to different bands. For the $2^+_1 \rightarrow 0^+_2$ transition (where the Hoyle
state is treated as a bound state) we obtain a strength of $\sim 0.9$ $e^2\mbox{fm}^4$, in 
reasonably good agreement with the result of the microscopic $\alpha$-cluster calculation 
given in Ref.\cite{che07},
but clearly smaller than the experimental value. For the $2^+_2 \rightarrow 0^+_1$ transition our
computed strengths are quite stable, although the agreement with the microscopic
$\alpha$-cluster calculation 
given in Ref.\cite{che07} is better, as expected, when the energy of the $2^+_2$ resonance is
put at about 2.4 MeV, also in better agreement with the experimental value. However, these 
energies of about 2 MeV are in clear disagreement with the AMD result. When compared to the experimental 
data, our strength
is about a factor of 3 bigger than the result given in Ref.\cite{zim13}, but in better agreement
with the value given in Ref.\cite{zim13b}, where the same experiment as in Ref.\cite{zim13} is
reexamined, see Fig.\ref{fig3}. For the $4^+_2 \rightarrow 2^+_2$ transition our results 
are reasonably consistent with the AMD calculation, especially when considering the 
${\cal B}_\gamma$ values.  The computed strengths for the transitions
$4^+_1 \rightarrow 2^+_1$ and $2^+_2 \rightarrow 2^+_1$ are also given, although for these cases 
the comparison with previous results or experimental data is not possible.

In general, the computed ${\cal B}_\sigma$ and ${\cal B}_\gamma$ strengths are consistent with
each other, especially if we take into account that the computed ${\cal B}_\sigma$ values
are obviously dependent on the width chosen for the energy windows, and the ${\cal B}_\gamma$ strengths are
very sensitive to the value of $E_\gamma$ used in Eq.(\ref{gamma}), since the photon energy
appears to the fifth power. This is particularly true when the initial and final energies are not
very far from each other, like for instance in the $2^+_2 \rightarrow 0^+_2$ transition with 
the $2^+_2$ energy
at 1.7 MeV. In this case a variation of $E_\gamma$ from 1.3 MeV to 1.4 MeV (and $\Gamma_\gamma=1$ meV)
implies a change in ${\cal B}_\gamma$ from 230 $e^2$fm$^4$ to 330 $e^2$fm$^4$. 
Together with this, the results involving the $4^+_1$ and $4^+_2$ states have the additional
uncertainty arising from the interference between them and the soft-photon contribution.

When comparing the results with the Ali-Bodmer and Buck potentials, the general conclusion is that
there are no significant differences between them. Only in the transitions involving the $2^+_2$ 
resonance at 1.7 MeV an important difference, mainly for the ${\cal B}_\gamma$ values, has been observed. This is due to the very
different width obtained for this resonance with each of the potentials. Therefore, from the
transition strengths it is not possible to answer the question of what potential is more 
appropriate in order to describe the alpha-alpha interaction. 

\subsection{Rotational model}

For transitions between states within a schematic rotational band, and assuming
axial symmetry for the system, the quadrupole transition strength is given by \cite{bohr}
\begin{equation}
{\cal B}^{(E2)}(J\rightarrow J')=\frac{5}{16\pi} Q_0^2
\langle J 0; 2 0|J' 0\rangle^2,
\label{eq19}
\end{equation}
where the projection, $K$, of the angular momentum on the intrinsic symmetry axis 
has been assumed to be zero.  The intrinsic quadrupole moment $Q_0$ is given by:
\begin{equation}
Q_0= \langle\sum_{i} q_i (2 z_i^2 - x_i^2 - y_i^2)\rangle 
   \; ,
\label{eq20}
\end{equation}
where $i$ runs over all the charged particles with charge $q_i$ and
whose center of mass coordinates are denoted by $(x_i,y_i,z_i)$, where 
the $z$-axis is
chosen along the intrinsic symmetry axis.  The expectation value
is taken in the intrinsic body-fixed coordinate system.
The quadrupole moment is a measure of the deformation and it has
ideally one characteristic value for a sequence of states belonging to
a given rotational band.  From each of the calculated transition
strengths given in Table~\ref{tab5}, it is then easy, to obtain 
from Eq.(\ref{eq19}) the absolute value of the intrinsic quadrupole moments.  
The $|Q_0|$-values obtained in this way are given in Table~\ref{tab6}.

\begin{table*}
\begin{center}
\caption{Absolute value of the intrinsic transition quadrupole moments $|Q_0|$ (in $e\mbox{fm}^2$) obtained from the ${\cal B}(E2)$ transition strengths given in Table~\ref{tab5} and Eq.(\ref{eq19}).
$|Q_0|_\sigma$ and $|Q_0|_\gamma$ denote the intrinsic quadrupole moments obtained from the transition strengths ${\cal B}_\sigma$ and ${\cal B}_\gamma$, respectively.  }
\label{tab6}
\begin{ruledtabular}
\begin{tabular}{|c|ccc|cccc|cccc|}
Transition  &  Exp. & $\alpha$-cluster &  AMD &\multicolumn{4}{c|}{Ali-Bodmer}  & \multicolumn{4}{c|}{Buck}  \\ \hline
$2^+_1 \rightarrow 0^+_1$ &  $19.5\pm0.5$& 21.46 & 20.5  &\multicolumn{4}{c|}{22.6} & \multicolumn{4}{c|}{22.3}  \\
$4^+_2 \rightarrow 2^+_1$ &           &      & 23.6  &\multicolumn{2}{c}{$|Q_0|_\gamma=17.8$}  & \multicolumn{2}{c|}{$|Q_0|_\sigma=17.0$} & \multicolumn{2}{c}{$|Q_0|_\gamma=21.0$}  &  \multicolumn{2}{c|}{$|Q_0|_\sigma=18.7$}  \\ \hline
   &   &    &    & \multicolumn{2}{c}{$E_{2^+}$=1.7 MeV} & \multicolumn{2}{c|}{$E_{2^+}$=2.4 MeV} &
             \multicolumn{2}{c}{$E_{2^+}$=1.7 MeV} & \multicolumn{2}{c|}{$E_{2^+}$=2.4 MeV}    \\
   &   &    &    &  $|Q_0|_\gamma$ &   $|Q_0|_\sigma$ &  $|Q_0|_\gamma$ &   $|Q_0|_\sigma$ &  $|Q_0|_\gamma$ &  $|Q_0|_\sigma$ &   $|Q_0|_\gamma$ &   $|Q_0|_\sigma$   \\
$2^+_2 \rightarrow 0^+_2$ &           &      & 72 &  108  & 94  & 92 & 84 &  78  &  82 & 89  & 83  \\
$4^+_1 \rightarrow 2^+_2$ &           &      & 145 & 77   & 69  & 73 & 68 &  73  &  68 & 88  & 71  \\ \hline
$2^+_1 \rightarrow 0^+_2$ &  $11.4\pm0.9$& 6.50 & 16.0  &\multicolumn{4}{c|}{6.8} & \multicolumn{4}{c|}{6.9}  \\
$4^+_1 \rightarrow 2^+_1$ &           &      &               &\multicolumn{2}{c}{$|Q_0|_\gamma=6.5$}  & \multicolumn{2}{c|}{$|Q_0|_\sigma=4.6$} & \multicolumn{2}{c}{$|Q_0|_\gamma=6.5$}  &  \multicolumn{2}{c|}{$|Q_0|_\sigma=5.9$}  \\ 
   &   &    &    & \multicolumn{2}{c}{$E_{2^+}$=1.7 MeV} & \multicolumn{2}{c|}{$E_{2^+}$=2.4 MeV} &
             \multicolumn{2}{c}{$E_{2^+}$=1.7 MeV} & \multicolumn{2}{c|}{$E_{2^+}$=2.4 MeV}     \\
   &   &    &    &  $|Q_0|_\gamma$ &   $|Q_0|_\sigma$ &  $|Q_0|_\gamma$ &   $|Q_0|_\sigma$ &  $|Q_0|_\gamma$ &  $|Q_0|_\sigma$ &   $|Q_0|_\gamma$ &   $|Q_0|_\sigma$   \\
$2^+_2 \rightarrow 0^+_1$ &$6.03\pm 0.54$& 10.0 & 4.5 & 14.3 & 12.5 & 10.0 & 9.8 & 11.0  & 11.9 & 10.8 & 10.5 \\
                          &$8.88\pm 0.37$&      &     &      &      &      &      &       &      &      &      \\
$2^+_2 \rightarrow 2^+_1$ &           &      &      &  11.2 & 10.1 & 13.9 & 13.5 & 9.4 & 10.1 &  13.9 & 13.8      \\
$4^+_2 \rightarrow 2^+_2$ &           &      & 16.2 & 17.3  & 11.9 & 18.6 & 13.3 & 15.1& 11.1 & 16.2 & 12.9    \\
\end{tabular}
\end{ruledtabular}
\end{center}
\end{table*}

The intrinsic quadrupole moment $Q_0$ is
related to the static quadrupole moment $Q$ of each
individual state with angular momentum $J$ by the simple expression:
\begin{equation}
Q=
\sqrt{\frac{16\pi}{5} } \langle J J |\hat{\cal O}_{20} | J J \rangle
=-\frac{J}{2J+3}Q_0,
\label{q0}
\end{equation}
where the operator $\hat{\cal O}_{\lambda\mu}$ is given in
Eq.(\ref{oper}) and, again, a band with $K=0$ has been assumed.

For the transitions between the states in the first band, $2^+_1
\rightarrow 0^+_1$ and $4^+_2 \rightarrow 2^+_1$ (the state $4^+_2$ is
the one assigned to the first band), the computed values
of $|Q_0|$ in Table~\ref{tab6} are rather stable, consistent with each
other, and consistent as well with the experimental value and previous
calculations.  The value of the static quadrupole moment 
$Q$ can be easily computed for the bound
$2^+_1$ state, and we have obtained 6.6 $e \mbox{fm}^2$ and 6.5 $e
\mbox{fm}^2$ with the Ali-Bodmer and Buck potentials, respectively.
These values agree with the experimental value of $6\pm 3$ $e
\mbox{fm}^2$ given in \cite{ver83}.  From Eq.(\ref{q0}) we can then
extract the intrinsic quadrupole moment for the $2^+_1$ bound state,
which is found to be $-22.7$ $ e \mbox{fm}^2$ and $-23.0$ $e
\mbox{fm}^2$ with the Ali-Bodmer and Buck potentials, respectively.
This result agrees as well with the $-21.6$ $e \mbox{fm}^2$ given in the
Ref.\cite{kam81}.  These values
are also consistent with the ones quoted in the
first two rows in Table~\ref{tab6} for the $2^+_1 \rightarrow 0^+_1$
and $4^+_2 \rightarrow 2^+_1$ transitions.  These results support the
idea of the states in this band as arising from the rotation of a
given intrinsic state, and also the fact the $4^+_2$ state is the
one actually belonging to the first band.

For the transitions in the second band, $2^+_2 \rightarrow 0^+_2$
and $4^+_1 \rightarrow 2^+_2$, the relative difference between the computed
values of $|Q_0|$ is higher than the one found in the first band. 
Typically, the results for the $4^+_1 \rightarrow 2^+_2$ transition are even a 25\% 
smaller than the ones for the $2^+_2 \rightarrow 0^+_2$ transition. If we restrict 
ourselves to the results involving the $2^+_2$ state at 2.4 MeV, which is in
better agreement with the experimental value, the computed $|Q_0|$ values 
range from $\sim 70$ $e \mbox{fm}^2$ up to $\sim 90$ $e \mbox{fm}^2$. 
Taking into account all the uncertainties already discussed, especially when 
the $4^+_1$ state is involved, we can consider that these results are consistent with 
each other. In this connection, it is important to remember that, as discussed
in Fig.\ref{fig7}b, the transition strengths obtained from the $4^+_1$ peak 
for the $4^+_1 \rightarrow 2^+_2$ reaction should be taken as its minimum value.
A small decrease in the photon energy cutoff would enhance the $4^+_1$ peak in
Fig.\ref{fig7}b, making the strengths of 
the $4^+_1 \rightarrow 2^+_2$ process closer to the ones of the $2^+_2 \rightarrow 0^+_2$
reaction. Therefore, we can conclude that the transition strengths given in 
Table~\ref{tab6} between the states in the second band in $^{12}$C are also consistent with 
the idea of a relatively well 'frozen' structure and the rotational character of the band. 
For the same reason,
we can also consider that our results are not dramatically far
from the 72 $e$fm$^2$ and 145 $e$fm$^2$ obtained in Ref.\cite{cuo13}.

In the lower part of Table~\ref{tab6} we give the intrinsic quadrupole
moments for the transitions between states in different bands. It is 
interesting to note that for the $4^+_1 \rightarrow 2^+_1$ transition,
which in principle should be a transition belonging to the first 
rotational band, we obtain a value for $|Q_0|$ even a factor of 3 smaller
than the values quoted in the upper part of the table for the 
$2^+_1 \rightarrow 0^+_1$ and $4^+_2 \rightarrow 2^+_1$ transitions.
This result is then also consistent with the assignment of the $4^+_2$
state to the first band and the $4^+_1$ state to the second. In 
fact, as also seen in the table, the $|Q_0|$ value for the 
$4^+_2 \rightarrow 2^+_2$ is also clearly smaller than the values 
for the transitions between the states in the second band.

\begin{figure}
\epsfig{file=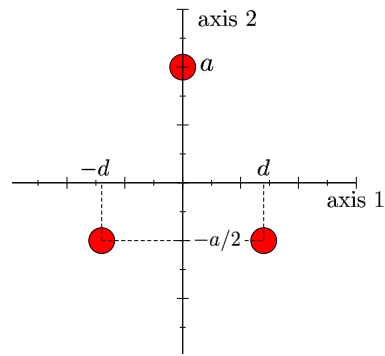,width=5.cm,angle=0}
\caption{(Color online)
Coordinates of the three-alpha system in the center of mass frame.  }
\label{fig10}
\end{figure}

A different approach can be made by applying the definition given in
Eq.(\ref{eq20}) to a system made of three point-like $\alpha$-particles
in a schematic triangular structure, as shown in Fig.\ref{fig10}. From the coordinates
given in the figure, we can obtain the hyperradius, $\rho_{3b}$, of such a three-body system,
which is given by:
\begin{equation}
\rho_{3b}^2=\frac{m_\alpha}{m} \frac{a^2}{2} \left(3+4(d/a)^2\right),
\label{eq23}
\end{equation}
where $m_\alpha$ is the mass of the $\alpha$-particle and $m$ is the normalization
mass used to define the Jacobi coordinates (the nucleon mass in our calculations).

Also, using the coordinates given in Fig.\ref{fig10}, we can get the moments of inertia relative 
to each of the three coordinate axes (where axis-3 is perpendicular to the plane
containing the three particles shown in the figure). These three moments of inertia 
take the form:
\begin{eqnarray}
{\cal I}_1 &=& \frac{3}{2} m_\alpha a^2 \label{i1}\\
{\cal I}_2 &=& 2 m_\alpha d^2 \label{i2}\\
{\cal I}_3 &=& {\cal I}_1+{\cal I}_2= \frac{3}{2} m_\alpha a^2 +2 m_\alpha d^2 \label{i3},
\end{eqnarray}
where in ${\cal I}_i$ the index $i$ refers the axis with respect to which the moment
of inertia has been calculated.

We have to keep in mind that in Eqs.(\ref{eq19}) and (\ref{eq20}) an axial symmetry has
been assumed. This implies that these equations can be used in those cases in which two of the 
moment of inertia in Eqs.(\ref{i1}) to (\ref{i3}) are equal to each other, while the
axis with respect to which the moment of inertia is different to the other two defines
the symmetry axis (and therefore the $z$-axis in Eq.(\ref{eq20})).  In other words, there 
are three possible geometries for the system in Fig.\ref{fig10} to which Eqs.(\ref{eq19}) 
and (\ref{eq20}) can be applied. They are the geometries corresponding to having the symmetry 
axis along axis-1, axis-2, or axis-3:

\begin{figure}
\epsfig{file=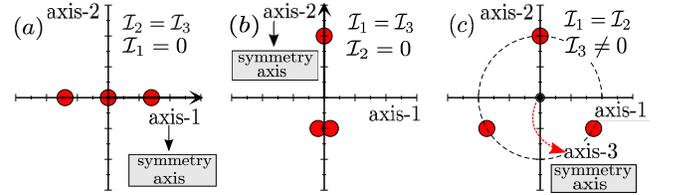,width=8.5cm,angle=0}
\caption{(Color online) Geometries in the three-alpha system in Fig.~\ref{fig10} corresponding
to the symmetry axis along axis-1 $(a)$, along axis-2 $(b)$, and along axis-3 $(c)$. The third axis,
axis-3, is perpendicular to the plane holding the three alphas. }
\label{fig11}
\end{figure}

$(a)$ Symmetry axis along axis-1 (${\cal I}_2={\cal I}_3$). Making equal Eqs.(\ref{i2}) and
(\ref{i3}) we get that this happens when $a=0$, leading to ${\cal I}_2={\cal I}_3=2m_\alpha d^2$ and
${\cal I}_1=0$. Making $a=0$ corresponds to the geometry shown
in Fig.\ref{fig11}a, where the three alphas are aligned along axis-1, which is the symmetry
axis. Therefore when computing $Q_0$ by use of Eq.(\ref{eq20}) the $z$-axis has to be
taken along axis-1, and for the three-alpha system ($q_i=2e$) we get $Q_0=8ed^2$, which by means
of Eq.(\ref{eq23}) and taking $m_\alpha/m\approx 4$ can be written as $Q_0=e \rho_{3b}^2$.

$(b)$ Symmetry axis along axis-2 (${\cal I}_1={\cal I}_3$). Making now equal Eqs.(\ref{i1}) and
(\ref{i3}) we get that this happens when $d=0$, and we get in this case 
${\cal I}_1={\cal I}_3=3m_\alpha a^2/2$ and ${\cal I}_2=0$. The corresponding geometry is now
the one shown in Fig.\ref{fig11}b, with the three particles along axis-2, one of them 
at the position $a$ and the other two at $-a/2$. Taking then axis-2 as the $z$-axis, we get 
from Eq.(\ref{eq20}) that for the three alphas $Q_0=6ea^2$, which making use of Eq.(\ref{eq23})
can again be written as $Q_0=e \rho_{3b}^2$.

$(c)$ Symmetry axis along axis-3 (${\cal I}_1={\cal I}_2$). As before, making equal 
Eqs.(\ref{i1}) and (\ref{i2}) we find that the symmetry axis is along axis-3 when 
$d^2/a^2=3/4$, and ${\cal I}_1={\cal I}_2=3m_\alpha a^2/2$, and ${\cal I}_3=3m_\alpha a^2$.
In this case the corresponding geometry is the one shown in Fig.\ref{fig11}c, which corresponds
to the three alphas in an equilateral triangle. Taking then the $z$-axis along axis-3 (perpendicular
to the plane holding the three alphas) we get from Eq.(\ref{eq20}) that $Q_0=-6ea^2$, or, using
again Eq.(\ref{eq23}), $Q_0=-e\rho_{3b}^2/2$.

Summarizing, the three axially symmetric geometries shown in Fig.\ref{fig11} lead to either
$Q_0=e\rho_{3b}^2$, which happens for the linear geometries $(a)$ and $(b)$, or 
$Q_0=-e\rho_{3b}^2/2$, which happens for the equilateral triangular geometry (c).

As already discussed, our three-body calculations are consistent with $Q_0\approx -23 \mbox{ $e$fm$^2$}$
for the bound $2^+_1$ state, as given for instance in \cite{ver83,kam81}. Such a negative value for the
intrinsic quadrupole moment is only consistent with $Q_0=-e\rho_ {3b}^2/2$, and therefore with the equilateral
structure. Furthermore, as shown in Table~\ref{tab3}, for the states in the first band we have that 
$\rho_{rms} \approx 6.7$ fm, from which we get $Q_0=-e\rho_{rms}^2/2\approx -22.5 \mbox{ 
$e$fm$^2$}$.  This result is 
in very good agreement with the computed intrinsic quadrupole moment for the $2^+_1$ state, the
values given in \cite{ver83,kam81}, and also with the values given in Table~\ref{tab6} for the
transitions between the states in the first band. We then conclude that the computed transition
strengths and intrinsic quadrupole moments for the transitions between the states in the first band
in the three-alpha system are consistent with the ones corresponding to transitions between the states
of a rotational band in a three-alpha system where the three alphas are sitting in the vertices of
an equilateral triangle.

For the states in the second band, we have from Table~\ref{tab3} that 
$\rho_{rms} \approx 10$ fm. Using this value we get that 
for the linear structures in Fig.\ref{fig11}a and Fig.{\ref{fig11}b the intrinsic
quadrupole moment in the band should be $Q_0\approx 100 \mbox{ $e$fm$^2$}$,
and for an equal sided triangular distribution (Fig.\ref{fig11}c) it should be
$Q_0\approx -50 \mbox{ $e$fm$^2$}$. Keeping in mind that, as discussed in 
connection with Fig.\ref{fig7}b, the results shown in Table~\ref{tab6}
for the $4^+_1 \rightarrow 2^+_2$ transition could have been underestimated,
we can conclude that our computed intrinsic quadrupole moments for the states in the
second band are consistent with the rotational estimate of 
$\sim 100 \mbox{ $e$fm$^2$}$ corresponding to the aligned structure shown in either 
Fig.\ref{fig11}a or Fig.\ref{fig11}b.

Another hint about the rotational character of the two bands in $^{12}$C can be
given by the sequence of energies in the $\left\{0^+_1, 2^+_1, 4^+_2\right\}$ 
and the $\left\{0^+_2, 2^+_2, 4^+_1\right\}$ bands. As it is well known,
in the case of axial symmetry and $K=0$ bands, these energies should follow the rule:
\begin{equation}
E_J-E_0=\frac{\hbar^2}{2{\cal I}} J(J+1),
\label{rota}
\end{equation}
where $E_J$ is the energy of the state in the band with angular momentum $J$, $E_0$
is the energy of the lowest state in the band, and ${\cal I}$ is the moment of
inertia relative to an axis perpendicular to the symmetry axis. 

For the three axially symmetric configurations given in Fig.\ref{fig11} we have:
\begin{eqnarray}
\mbox{Geometry $(a)$:\hspace*{5mm}} {\cal I}&=&2m_\alpha d^2=m_\alpha \rho_{3b}^2/4  \label{eq27}\\
\mbox{Geometry $(b)$:\hspace*{5mm}} {\cal I}&=&3m_\alpha a^2/2=m_\alpha \rho_{3b}^2/4 \label{eq28}\\
\mbox{Geometry $(c)$:\hspace*{5mm}} {\cal I}&=&3m_\alpha a^2/2=m_\alpha \rho_{3b}^2/8 \label{eq29},
\end{eqnarray}
where, again, we have made use of Eq.(\ref{eq23}) with $a=0$ in case $(a)$, $d=0$ 
in case $(b)$, and $d^2/a^2=3/4$ in case $(c)$.

For the first band, for which $E_0=-7.28$ MeV, the experimental energies of the $2^+_1$ and $4^+_2$
states given in Table~\ref{tab2} lead by means of Eq.(\ref{rota}) to $\hbar^2/2{\cal I} \approx 0.74$ MeV
and 0.71 MeV, respectively. These values are very similar to each other, supporting the 
fact that the first band fulfills the condition of a ``frozen'' structure that
gives rise to a rotational band. Furthermore, this first band has been seen to be
consistent with the equal sided triangular structure and $\rho_{rms}
\approx 6.7$ fm. Making use of Eq.(\ref{eq29}) we estimate from the rotational model
that $\hbar^2/2{\cal I} \approx 0.9$ MeV, which is relatively close to the
values obtained from the experimental $2^+_1$ and $4^+_1$ energies.

For the second band, for which $E_0=0.38$ MeV, the energies of the $2^+_2$ and $4^+_1$
states in Table~\ref{tab2} determine $\hbar^2/2{\cal I} \approx 0.33$ MeV and
0.25 MeV for the $2^+_2 \rightarrow 0^+_2$ and $4^+_1 \rightarrow 2^+_2$ transitions.  
From the analysis of the intrinsic quadrupole moments we concluded that the states
in the second band could at best be consistent with the linear structure in Figs.\ref{fig11}a 
and \ref{fig11}b, which by use of Eq.(\ref{eq27}) or (\ref{eq28}), and taking
$\rho_{rms}\approx 10$ fm (as shown in Table~\ref{tab3}), lead
to the estimate from the rotational model $\hbar^2/2{\cal I} \approx 0.2$ MeV, which
is again reasonably consistent with the values obtained from the energy differences. 
Thus, also in this case the analysis of the energies in the $\{0^+_2,2^+_2,4^+_1\}$ band
supports the conclusion obtained from the intrinsic quadrupole moments, namely,
the properties of the states in the second band of the three-alpha system are 
consistent with the behaviour expected for states belonging to a rotational band
whose structure would be the one in Fig.\ref{fig11}a or Fig.\ref{fig11}b.

Although the analysis of both, $Q_0$ and ${\cal I}$, 
does not permit to distinguish between the linear structures shown in Figs.\ref{fig11}a
and \ref{fig11}b, it is important to note that the alpha-alpha Coulomb repulsion clearly 
prevents the formation of a system as the one described in Fig.\ref{fig11}b. For this reason, 
together with the fact that the geometry given in Fig.\ref{fig11}b
is not consistent with cluster model calculations, 
we can conclude that the only possible aligned structure for the states in the
second band in $^{12}$C is the one given in Fig.\ref{fig11}a.

We emphasize that our rotational model analysis is schematic, even
within our $\alpha$-particle model. The $0^+$ and $2^+$ bound states
are known only to be approximately described by an $\alpha$-particle
model.  The continuum states investigated in the present paper are
on the other hand expected to be better $\alpha$-particle
states. However, this does not imply that a schematic classical linear
or triangular structure provide an accurate description. Already the
quantum mechanical probability distributions differ from this
simplified picture.  Both an equal sided triangle and a linear chain
can only reveal the qualitative essence of the structure of these
states.  In particular, the accuracy of this illuminating qualitative
picture is not in conflict with the bent arm structure of the Hoyle
state found in many previous calculations as well as in the present
one.

The present interpretation of a rotational band structure is at first
glance in conflict with the interpretations of both the $0^+_2$ and $2^+_2$
states as a very dilute gas of three particles \cite{fun03,fun05}.
These two interpretations are difficult to reconcile, since a gas of
particles does not have rotational states as defined in classical
textbooks \cite{sie87} and used in the present paper. Also, the
radii of these states in \cite{fun03,fun05} differ apparently from each
other, although ascribed to different energies but with ``similar
structure''. In any case, our conclusions about energies, transition
probabilities, and quadrupole moments rely entirely on the computed wave
functions.  In this sense, interpretations do not play any role, since they 
do not enter into any of the quantum mechanical solutions.
Whether these solutions are understood in terms of a frozen
intrinsic structure or a dilute gas-like structure is not important,
and both  interpretations can in principle be correct simultaneously.

\section{Summary and Conclusions}

In this work we investigate the spectrum of $^{12}$C and its
rotational character.  The existence of two $0^+$, $2^+$, $4^+$
sequences, related to the $0^+_1$ ground state and the $0^+_2$
resonance (the Hoyle state), led people to refer to these two series
as rotational bands in $^{12}$C.  The sequence, $0^+$, $2^+$, $4^+$,
suggests that the geometry corresponds to axial and $R_2$ symmetric
intrinsic states.  However, transitions between rotational states
build on different intrinsic shapes with $K=0$ bandheads would also be
mathematically described in precisely the same way.  This means that
the same tests apply even if other states with different quantum
numbers are members of the same band.

In the present case we investigated whether the properties of these
continuum states are consistent with the electromagnetic transition
strengths for a rotational band built on the ground and first excited
$0^+$ states.  In particular, the transition strength should equal the
square of the intrinsic quadrupole moment multiplied by a
Clebsch-Gordan coefficient.

The three-body wave functions in $^{12}$C have been obtained through
the hyperspherical adiabatic expansion method, and the continuum
spectrum has been discretized by imposing a box boundary condition on
the hyperradius.  The discretized continuum spectrum has been obtained
on the real energy axis, without preferential treatment of the
resonances.  Thus, we have from the outset the natural distribution
over continuum states. Most existing other models treat resonances as
bound sates even though many of them have a substantial width.  All
transition strengths are then present in one wave bound state like
resonance function.

The transition strength for the different reactions has been extracted
from the corresponding $\gamma$-emission cross sections. For
transitions into a resonance, the calculation of the cross section as
a function of the initial three-body energy requires specification of
the range of energies investigated in the final state, which should be
around the final resonance energy. For this reason, some information
about the position of the resonances is needed. In this work this is
done by means of the complex scaling method, which provides the
resonances as poles of the ${\cal S}$-matrix.  It is important to
emphasize that this is done only to know what energy windows to
consider for transitions into a resonance, since the method used to
discretize the continuum and to compute the cross sections only deals
with discretized continuum states with real energies.

From the cross sections two different methods have been used to obtain the transition 
strengths. The first one uses the $\Gamma_\gamma$-width as a parameter to fit the peaks of
the cross section, associated to resonances in the initial state, with the usual Breit-Wigner
shape. From the $\Gamma_\gamma$-width the transition strength is computed making use of the simple
expression relating these two quantities. In the second method the cross section, as a function
of the incident energy, is integrated under the peak corresponding to initial resonance. This 
integrated cross section gives, except for some known factors, the transition strength.

The cross sections for the $4^+$ to $2^+$, the $2^+$ to $0^+$, and $2^+$ to $2^+$ 
transitions have then been computed. 
Two different potentials, the Ali-Bodmer and the Buck potentials, have been used to describe
the $\alpha$-$\alpha$ interaction. The main features of the computed cross sections are in general
independent of the potential used. Only those reactions involving resonances whose properties are more
sensitive to the potential show a sizable difference in the cross section. This happens
for instance in the reactions with initial states with angular momentum and parity $2^+$, 
especially when the lowest $2^+$ resonance is located at about 1.7 MeV, for which the predicted
resonance width differs by a factor of 2 depending on the potential used. In any case, when
the energy of the $2^+_2$ resonance is moved up to 2.4 MeV (in better agreement with the recent
experimental value) the dependence on the potential is much less relevant.

The transition strengths obtained from the cross sections are in general consistent 
with each other. First, the results are similar no matter what method is used to extract the 
transition strengths (${\cal B}_\sigma$ or ${\cal B}_\gamma$), especially when taking into account
the uncertainties inherent to each of the methods. And, second, they are also independent
of the potential used. Again, only in the reactions involving the $2^+_2$ state at 1.7 MeV
show a higher dependence on the potential.

The first result we obtained is that the $4^+_1$ and $4^+_2$ states, which belong to 
different bands in $^{12}$C have actually crossed, in such a way that the $4^+_1$ state 
belongs to the second band and the $4^+_2$ to the first one. This is first suggested by
the fact that the computed values of $\rho_{rms}$, which should be similar for 
systems having the same 'frozen' spatial structure, provides values for the $4^+_1$ state
similar to the ones of the $0^+_2$ and $2^+_2$ states, and values for the $4^+_2$ state
similar to the ones of the $0^+_1$ and $2^+_1$ states. Furthermore, investigating how the
$4^+_1$ and $4^+_2$ resonances move in the complex energy plane when making the effective
three-body force more and more attractive, we have seen that the first $4^+$ state becoming
bound would actually be the $4^+_2$.  

The assignment of the $4^+_1$ state to the second band and the $4^+_2$ to the first 
is also confirmed by the analysis of the transitions strengths and the intrinsic 
quadrupole moments. The values of $Q_0$ are obtained assuming a rotational character
for the two bands in $^{12}$C under investigation, in such a way that the transition
strength is basically the square of the intrinsic quadrupole moment multiplied by some
geometrical factor depending on the initial and final angular momenta of the transition.
Doing like this we have seen that for the $4^+_2 \rightarrow 2^+_1$ 
reaction the computed transition strength is clearly more consistent with the previous 
AMD calculation than the one corresponding to the $4^+_1 \rightarrow 2^+_1$. Furthermore, 
the intrinsic quadrupole
moments derived from the transition strengths for the $2^+_1 \rightarrow 0^+_1$ and
 $4^+_2 \rightarrow 2^+_1$ reactions are rather similar and consistent with previous
calculations and the experimental value. This consistency would disappear if the 
reaction $4^+_1 \rightarrow 2^+_1$ were the one taken into account. The same happens
for the reactions in the second band. Although the computed transition strength for 
the $4^+_1 \rightarrow 2^+_2$ is even a factor of 3 smaller than the previous 
AMD calculation, this discrepancy is certainly less important than when considering
the $4^+_2 \rightarrow 2^+_2$ reaction, and it can actually be understood from the
uncertainties associated to the method used to extract the strength. Also, the 
intrinsic quadrupole moments for the $2^+_2 \rightarrow 0^+_2$ and $4^+_1 \rightarrow 2^+_2$ 
are reasonably stable, stability
that would be clearly broken if assuming the $4^+_2 \rightarrow 2^+_2$ transition
as belonging to the second band. Therefore, from the stability of the intrinsic
quadrupole moments for each of the $\{0^+_1,2^+_1,4^+_2\}$ and $\{0^+_2,2^+_2,4^+_1\}$
bands we can conclude that each of the bands correspond to states having a rather
well preserved rigid structure.

The rotational character of two bands is confirmed when comparing with the prediction
obtained from an axially symmetric rotating structure made by three point-like alpha
particles. Assuming values for $\rho_{3b}$ similar to the ones computed numerically for
the states in the first and second bands ($\rho_{rms}\sim 6.7$ fm and $\sim 10$ fm, respectively)
we have seen that the $Q_0$ values previously computed for the states in the first band
are consistent with a equilateral triangular structure rotating around an axis perpendicular
to the plane holding the three particles. For the states in the second band, the computed
$Q_0$ values are mostly consistent with a linear distribution of the particles. The same results
are obtained from the analyses of the moments of inertia.

Summing up, we have calculated genuine continuum-continuum electric
quadrupole ($E2$) transitions from $0^+$, $2^+$ and $4^+$ states in
$~{12}$C.  The transition strengths are not well defined since neither
initial nor final states have well defined bound-state like structures
in the continuum.  Energy windows around resonance positions extending
at least the natural widths must contribute to the transitions.
Instead we design, approximate and deduce correspondingly
non-observable transition strengths from (double) differential cross
sections.  The results are two sequences of rotational band
structures with ground and Hoyle states as band heads with interchange
of the order of the two close-lying $4^+$-resonances.  The derived
intrinsic quadrupole moments and moments of inertia are consistent with
axial triangle and almost linear structures, respectively.  In
conclusion, our procedure to compute continuum-continuum transitions
are used to classify six $~{12}$C-states in two rotational bands.

\appendix
\section{Three-body incoming flux}
\label{apen1}

Given the three-body reaction $a+b+c \rightarrow A + \gamma$, the cross section corresponding to this
process requires a definition of the incoming flux of particles. This can be done similarly to the 
two-body case, where the incoming flux is well defined.

At the two body level the Schr\"{o}dinger equation takes the form:
\begin{equation}
i\hbar \frac{\partial \Psi}{\partial t}=
-\frac{\hbar^2}{2\mu} \bm{\nabla}_{\bm{r}}^2 \Psi + V \Psi,
\label{ap1}
\end{equation}
where $\mu$ is the reduced mass, $\Psi$ the two-body wave function, $\bm{\nabla}_{\bm{r}}$ is the
three-dimensional gradient operator in terms of the relative vector $\bm{r}$, and $V$ is the 
two-body potential. As described in any Quantum Mechanics text book, from the Schr\"{o}dinger 
equation we can obtain the flux of incoming particles as:
\begin{equation}
\bm{j}=\frac{1}{2\mu i}\left[ \Psi^* \bm{\nabla}_{\bm{r}}\Psi - (\bm{\nabla}_{\bm{r}}\Psi^*) \Psi\right],
\end{equation}
which for an incoming plane wave leads to the well-known result 
\begin{equation}
j=\hbar p/\mu, 
\label{two}
\end{equation}
with $p$ being the two-body relative momentum.

At the three-body level, the Schr\"{o}dinger reads:
\begin{equation}
i\hbar \frac{\partial \Psi}{\partial t}=
-\frac{\hbar^2}{2\mu_x} \bm{\nabla}_{\bm{r}_x}^2 \Psi 
-\frac{\hbar^2}{2\mu_y} \bm{\nabla}_{\bm{r}_y}^2 \Psi + V \Psi,
\label{ap3}
\end{equation}
where now $\mu_x$ is the reduced mass of two of the particles, $\bm{\nabla}_{\bm{r}_x}$ is
the gradient operator in terms of the relative coordinate $\bm{r}_x$ between those two particles,
$\mu_y$ is the reduced mass of the third particle and the two-body system made by the other two, and
$\bm{\nabla}_{\bm{r}_y}$ is the gradient operator in terms of the relative coordinate $\bm{r}_y$
between the third particle and the center of mass of the other two. The wave function $\Psi$ is now
a three-body wave function, and $V$ contains all the potentials involved in the three-body system.

Let us introduce now the usual Jacobi coordinates $\bm{x}=\sqrt{\mu_x/m}\bm{r}_x$ and
$\bm{y}=\sqrt{\mu_y/m}\bm{r}_y$ \cite{nie01}, which contain the arbitrary normalization mass $m$.
From this definition we immediately get that 
$\bm{\nabla}_{\bm{r}_x}=\sqrt{\mu_x/m} \bm{\nabla}_{\bm{x}}$
and $\bm{\nabla}_{\bm{r}_y}=\sqrt{\mu_y/m} \bm{\nabla}_{\bm{y}}$, where 
$\bm{\nabla}_{\bm{x}}$ and $\bm{\nabla}_{\bm{y}}$ are the gradient operators in terms of the
$\bm{x}$ and $\bm{y}$ Jacobi coordinates, respectively, and which permit to write the Schr\"{o}dinger
equation Eq.(\ref{ap3}) as
\begin{equation}
i\hbar \frac{\partial \Psi}{\partial t}=
-\frac{\hbar^2}{2m} \bm{\nabla}_{\bm{x},\bm{y}}^2 \Psi + V \Psi,
\label{ap4}
\end{equation}
which is equivalent to Eq.(\ref{ap1}) and where we have introduced
$\bm{\nabla}_{\bm{x},\bm{y}}^2=\bm{\nabla}_{\bm{x}}^2+\bm{\nabla}_{\bm{y}}^2$

Therefore, from Eq.(\ref{ap4}), and proceeding exactly as done at the two-body level, we can define
the three-body incoming flux as:
\begin{equation}
\bm{j}=\frac{1}{2 m i}\left[ \Psi^* \bm{\nabla}_{\bm{x},\bm{y}}\Psi - 
(\bm{\nabla}_{\bm{x},\bm{y}}\Psi^*) \Psi\right],
\label{ap6}
\end{equation}
where $\bm{\nabla}_{\bm{x},\bm{y}}$ is the six-dimensional gradient operator in the Jacobi coordinates
$\bm{x}$ and $\bm{y}$.

Considering now the incoming plane wave $e^{i(\bm{x}\cdot\bm{k}_x+\bm{y}\cdot\bm{k}_y)}$ we get
the incoming flux:
\begin{equation}
\bm{j}=\hbar \frac{(\bm{k}_x,\bm{k}_y)}{m},
\end{equation}
where $(\bm{k}_x,\bm{k}_y)$ represents the six-dimensional momentum given by the three-body momenta
$\bm{k}_x$ and $\bm{k}_y$, which are the momenta associated to the Jacobi coordinates $\bm{x}$
and $\bm{y}$. If we define now $\kappa=\sqrt{k_x^2+k_y^2}$ we then get:
\begin{equation}
j=\hbar \frac{\kappa}{m},
\end{equation}
which is completely analogous to the two-body result in Eq.(\ref{two}).

The result given above is correct provided that the wave function $\Psi$ is normalized to 1 in terms of
the Jacobi coordinates $\bm{x}$ and $\bm{y}$. However, the correct normalization should be done not in 
terms of the Jacobi coordinates, but in terms of the relative coordinates $\bm{r}_x$ and $\bm{r}_y$. 
Since
\begin{equation}
\int d\bm{x} d\bm{y} |\Psi|^2=\left(\frac{\mu_x}{m} \right)^{3/2}
              \left(\frac{\mu_y}{m} \right)^{3/2} \int d\bm{r}_x d\bm{r}_y |\Psi|^2,
\end{equation}
we have that to get $\Psi$ normalized to 1 in terms of the relative coordinates we have to multiply
$\Psi$ by the factor $(m/\mu_x)^{3/4}(m/\mu_y)^{3/4}$, that when done in Eq.(\ref{ap6}) 
leads to the final expression for the incoming flux:
\begin{equation}
j=\hbar \frac{\kappa}{m}
  \left(\frac{m}{\mu_x} \right)^{3/2} \left(\frac{m}{\mu_y} \right)^{3/2}. 
\label{ap10}
\end{equation}

It is important to note that the flux of particles given above depends on the arbitrary normalization 
mass $m$. This is reflecting the fact that for a given value of the hyperradius $\rho=\sqrt{x^2+y^2}$
the relative distances between the three incoming particles, $r_x$ and $r_y$, will be different for 
different values of the normalization mass. It is then clear than the flux of particles through some 
given hypersurface
has to depend as well on the choice of $m$. The cross section for the reaction 
$a+b+c \rightarrow A+\gamma$ is just the flux of outgoing particles normalized with the incoming flux,
and therefore, according to Eq.(\ref{ap10}), the cross section will depend as well on $m$. 
In other words, the cross section is only well defined for a given definition of the hyperradius.

\section{Cross section for the $a+b+c \rightarrow A + \gamma$ process}
\label{apen2}

As discussed above, the cross section for the process $a+b+c \rightarrow A+\gamma$ 
depends on the normalization mass $m$. It is not a well defined physical observable. In fact, 
the usual observable (therefore independent of $m$) for this kind of reactions
is the reaction rate, which after multiplication by the density of particles gives the
number of reactions per unit time and unit volume in some specific environment. The reaction rate 
for the $a+b+c \rightarrow A+\gamma$ process at a given three-body kinetic energy $E$, $R_{abc}(E)$, 
is given in Eq.(1) of Ref.\cite{gar11}:
\begin{eqnarray}
\lefteqn{
R_{a b c}(E)=}
 \label{apb1}\\ & &
\nu!\; \frac{\hbar^3}{c^2} \frac{8\pi}{(\mu_x \mu_y)^{3/2}} 
\frac{2(2 J_A+1)}{(2 J_a+1) (2 J_b+1) (2 J_c+1) }
\left( \frac{E_\gamma}{E} \right)^2 
\sigma_\gamma(E_\gamma),
\nonumber
\end{eqnarray}
where $\nu$ is the number of identical particles in the three-body system, $J_a$, $J_b$, and $J_c$
are the angular momenta of particles $a$, $b$, and $c$, $E_\gamma$ is the photon energy, and
$\sigma_\gamma(E_\gamma)$ is the photo-dissociation cross section of the
process $A+\gamma \rightarrow a+b+c$.

The reaction rate is given by the product of the cross section and the incoming flux. Using
the expression in Eq.(\ref{ap10}) we can then write:
\begin{equation}
R_{abc}(E)= \hbar \frac{\kappa}{m}
  \left(\frac{m}{\mu_x} \right)^{3/2} \left(\frac{m}{\mu_y} \right)^{3/2}
  \sigma_{abc}(E),
\end{equation}
which after use of Eq.(\ref{apb1}) leads to the following relation between the cross section 
for the process $a+b+c \rightarrow A + \gamma$ and the one corresponding
to the inverse reaction:
\begin{equation}
\frac{\sigma_{abc}(E)}{\sigma_\gamma(E_\gamma)}=
\nu! \frac{2 (2J_A+1)}{(2J_a+1)(2J_b+1)(2J_c+1)} 
\frac{32 \pi}{\kappa^5} \left(\frac{E_\gamma}{\hbar c} \right)^2 ,
\end{equation}
which depends on the normalization mass $m$ through $\kappa$.

\section{Reduced matrix element}
\label{apen3}

In this work the three-body wave functions are written as given in Eq.(\ref{eq7}):
\begin{equation}
\Phi_J={1\over\rho^{5/2}} \sum_n f^J_n(\rho) \phi_n^J(\rho,\Omega).
\label{ap31}
\end{equation}

The angular wave functions $\phi_n^J(\rho,\Omega)$ in Eq.(\ref{ap31}) are expanded
in terms of the hyperspherical harmonics:
\begin{equation}
\phi_n^J(\rho,\Omega)=\sum_q C_q^{(n)}(\rho) \left[{\cal Y}_{K \ell_x \ell_y}^L(\Omega) \otimes 
\chi_{s_x s_y}^S\right]^J,
\label{ap32}
\end{equation}
where $q$ collects the quantum numbers $\{K,\ell_x,\ell_y,L,s_x,S\}$, where $K$ is the hypermomentum,
$\ell_x$ and $s_x$ are the relative angular momentum and total spin of the two particles connected
by the $\bm{x}$ Jacobi coordinate, $\ell_y$ is the relative momentum between the third particle and the
center of mass of the first two, $L$ is the total angular momentum obtained by coupling
of $\ell_x$ and $\ell_y$, and $S$ is the total spin obtained by coupling $s_x$ and the
spin of the third particle $s_y$. The total angular momentum $J$ is obtained by coupling of $L$ and $S$.
The total spin function of the three-body system is denoted by $\chi_{s_x s_y}^S$, and the
hyperspherical harmonics ${\cal Y}_{K \ell_x \ell_y}^L(\Omega)$ are defined as:
\begin{eqnarray}
{\cal Y}_{K \ell_x \ell_y}^{LM_L}(\Omega)&=&N_K^{\ell_x \ell_y} (\sin \alpha)^{\ell_x}
(\cos \alpha)^{\ell_y} P_\nu^{\ell_x+1/2,\ell_y+1/2}(\cos 2\alpha) \nonumber \\ & & \times
\left[Y_{\ell_x}(\Omega_x)\otimes Y_{\ell_y}(\Omega_y)\right]^{LM_L},
\label{ap33}
\end{eqnarray}
where $\nu$ is such that $K=2\nu+\ell_x+\ell_y$ and the normalization constant $N_K^{\ell_x \ell_y}$
can be found for instance in \cite{nie01}.

For obvious reasons, in order to compute the matrix element 
$\langle \Phi_J^{(i)} || y_p^\lambda Y_{\lambda}(\hat{r}_p) || \Phi_{J'}^{(j)} \rangle$
it is convenient to write the three-body wave function in terms of the Jacobi set 
$\{\bm{x}_p, \bm{y}_p\}$. When this is done, insertion 
of Eqs.(\ref{ap33}) and (\ref{ap32}) into Eq.(\ref{ap31}) gives the full expansion
of the three-body wave function. Using this expression the integration over $\Omega_x$ and 
$\Omega_y$ involved in the calculation of the matrix element 
$\langle \Phi_{J M}^{(i)} | y_p^\lambda Y_{\lambda \mu }(\hat{r}_p) | \Phi_{J' M'}^{(j)} \rangle$
can be trivially made, and only the integrals over $\rho$ and $\alpha$ remain. 

After some algebra, we get the final result:
\begin{widetext}
\begin{eqnarray}
\langle \Phi_J^{(i)} || y_p^\lambda Y_{\lambda}(\hat{r}_p) || \Phi_{J'}^{(j)} \rangle  &=& 
\int d\rho \rho^\lambda \sum_n f^J_n(\rho) \sum_q C_q^{(n)}(\rho) N_K^{\ell_x\ell_y}
\sum_{n'} f_{n'}^{J'}(\rho)\sum_{q'}  C_{q'}^{(n')}(\rho)  N_{K'}^{\ell_x\ell'_y}
\frac{\hat{\lambda} \hat{\ell}_y \hat{\ell'}_y \hat{L} \hat{L'} \hat{J} \hat{J'}}{\sqrt{4\pi}}
\left(
    \begin{array}{ccc}
       \ell_y & \lambda & \ell'_y \\
       0 & 0 & 0
    \end{array}
    \right)  \nonumber \\ & &  \times
\left\{
    \begin{array}{ccc}
       L & \lambda & L' \\
       \ell'_y & \ell_x & \ell_y
    \end{array}
    \right\}
\left\{
    \begin{array}{ccc}
       J & J' & \lambda \\
       L' & L & S
    \end{array}
    \right\}
\frac{(-1)^{J'+\lambda+\ell_x+S}}{2^{a+b+2}} {\cal I}_\lambda^{\ell_x\ell_y \ell'_y K K'}
\delta_{\ell_x\ell'_x} \delta_{s_x,s'_x}\delta_{S S'},
\label{ap34}
\end{eqnarray}
\end{widetext}
where the indeces without and with primes refer to the $\Phi_J^{(i)}$ and $\Phi_{J'}^{(j)}$
wave functions, respectively, $\hat{u}=\sqrt{2u+1}$, $a=\ell_x+1/2$, and 
$b=(\lambda+1+\ell_y+\ell'_y)/2)$. Finally, ${\cal I}_\lambda^{\ell_x\ell_y \ell'_y K K'}$
denotes the integral over the hyperangle $\alpha$, which can be written as:
\begin{eqnarray}
\lefteqn{
{\cal I}_\lambda^{\ell_x\ell_y \ell'_y K K'}=} \\ & &
\int_{-1}^1 dx
(1-x)^a (1+x)^b
P_\nu^{(a,\ell_y+1/2)}(x) P_{\nu'}^{(a,\ell'_y+1/2)}(x).
\nonumber
\end{eqnarray}

\acknowledgments This work was partly supported by funds provided by
DGI of MINECO (Spain) under contract No. FIS2011-23565.  We appreciate
valuable continuous discussions with Drs. H. Fynbo and K. Riisager.

\end{document}